\documentclass[twocolumn,trackchanges]{aastex631}

\newcommand\Msol{\mbox{M$_\odot$}}
\newcommand{\tdep}{\mbox{$\tau_{\rm dep,mol}$}}

\newcommand{\sigsfr}{\mbox{$\Sigma_{\rm SFR}$}}
\newcommand{\sigstar}{\mbox{$\Sigma_{*}$}}

\received{September 21, 2023}
\accepted{January 19, 2024}
\submitjournal{ApJS}

\shorttitle{EDGE-CALIFA Survey Database}
\shortauthors{Wong et al.}
\newcommand{\kms}{$\mathrm{km~s^{-1}}$}
\newcommand\ttco{\mbox{$^{13}$CO}}
\newcommand\HI{\ion{H}{1}}
\newcommand\HII{\ion{H}{2}}
\newcommand\NII{\mbox{\ion{N}{2}}}
\newcommand\OIII{\mbox{\ion{O}{3}}}

\begin{document}

\title{The EDGE-CALIFA Survey: An Extragalactic Database for Galaxy Evolution Studies}

\correspondingauthor{Tony Wong}
\email{wongt@illinois.edu}

\author[0000-0002-7759-0585]{Tony Wong}
\affiliation{Department of Astronomy, University of Illinois, Urbana, IL 61801, USA}

\author[0000-0001-5301-1326]{Yixian Cao}
\affiliation{Max-Planck-Institut f\"ur extraterrestrische Physik, Giessenbachstra{\ss}e 1, D-85748 Garching, Germany}
\affiliation{Department of Astronomy, University of Illinois, Urbana, IL 61801, USA}

\author[0000-0002-4623-0683]{Yufeng Luo}
\affiliation{Department of Physics and Astronomy, University of Wyoming, Laramie, WY 82071, USA}

\author{Alberto D. Bolatto}
\affiliation{Department of Astronomy, University of Maryland, College Park, MD 20742, USA}

\author[0000-0001-6444-9307]{Sebasti\'{a}n F. S\'{a}nchez}
\affiliation{Instituto de Astronom\'{i}a, Universidad Nacional Aut\'{o}noma de M\'{e}xico, A.P. 70-264, 04510 M\'{e}xico, D.F.,  Mexico}

\author{Jorge K. Barrera-Ballesteros}
\affiliation{Instituto de Astronom\'{i}a, Universidad Nacional Aut\'{o}noma de M\'{e}xico, A.P. 70-264, 04510 M\'{e}xico, D.F.,  Mexico}

\author{Leo Blitz}
\affiliation{Department of Astronomy, University of California, Berkeley, CA 94720, USA}

\author{Dario Colombo}
\affiliation{Max-Planck-Institut f\"{u}r Radioastronomie, D-53121, Bonn, Germany}

\author{Helmut Dannerbauer}
\affiliation{Instituto de Astrof\'isica de Canarias, E-38205 La Laguna, Tenerife, Spain}
\affiliation{Universidad de La Laguna, Dpto.\ Astrof\'isica, E-38206 La Laguna, Tenerife, Spain}

\author{Alex Green}
\affiliation{Department of Astronomy, University of Illinois, Urbana, IL 61801, USA}

\author{Veselina Kalinova}
\affiliation{Max-Planck-Institut f\"{u}r Radioastronomie, D-53121, Bonn, Germany}

\author{Ferzem Khan}
\affiliation{Department of Astronomy, University of Illinois, Urbana, IL 61801, USA}

\author{Andrew Kim}
\affiliation{Department of Astronomy, University of Illinois, Urbana, IL 61801, USA}

\author{Eduardo A. D. Lacerda}
\affiliation{Instituto de Astronom\'{i}a, Universidad Nacional Aut\'{o}noma de M\'{e}xico, A.P. 70-264, 04510 M\'{e}xico, D.F.,  Mexico}

\author{Adam K. Leroy}
\affiliation{Department of Astronomy, The Ohio State University, Columbus, OH 43210, USA}

\author[0000-0003-2508-2586]{Rebecca C. Levy}
\altaffiliation{NSF Astronomy and Astrophysics Postdoctoral Fellow}
\affiliation{Steward Observatory, University of Arizona, Tucson, AZ 85721, USA}
\affiliation{Department of Astronomy, University of Maryland, College Park, MD 20742, USA}

\author[0000-0001-9068-6787]{Xincheng Lin}
\affiliation{Department of Astronomy, University of Illinois, Urbana, IL 61801, USA}

\author[0000-0002-0696-6952]{Yuanze Luo}
\affiliation{Department of Astronomy, University of Illinois, Urbana, IL 61801, USA}
\affiliation{William H. Miller III Department of Physics and Astronomy, Johns Hopkins University, Baltimore, MD 21218, USA}

\author[0000-0002-5204-2259]{Erik W. Rosolowsky}
\affiliation{Department of Physics, University of Alberta, Edmonton, T6G 2E1, Canada}

\author[0000-0002-5307-5941]{M\'onica Rubio}
\affiliation{Departamento de Astronom\'ia, Universidad de Chile, Casilla 36-D, Santiago, Chile}

\author[0000-0003-1774-3436]{Peter Teuben}
\affiliation{Department of Astronomy, University of Maryland, College Park, MD 20742, USA}

\author{Dyas Utomo}
\affiliation{Department of Astronomy, The Ohio State University, Columbus, OH 43210, USA}

\author[0000-0002-5877-379X]{Vicente Villanueva}
\affiliation{Department of Astronomy, University of Maryland, College Park, MD 20742, USA}

\author[0000-0002-8765-7915]{Stuart N. Vogel}
\affiliation{Department of Astronomy, University of Maryland, College Park, MD 20742, USA}

\author{Xinyu Wang}
\affiliation{Department of Astronomy, University of Illinois, Urbana, IL 61801, USA}

\begin{abstract}

The EDGE-CALIFA survey {provides spatially resolved optical} integral field unit (IFU) and CO spectroscopy for 125 galaxies selected from the CALIFA Data Release 3 sample.  {The Extragalactic Database for Galaxy Evolution (EDGE)} presents the spatially resolved products of the survey as pixel tables that reduce the oversampling in the original images and facilitate comparison of pixels from different images.  By joining these pixel tables to lower dimensional tables that provide radial profiles, integrated spectra, or global properties, it is possible to investigate the dependence of local conditions on large-scale properties.  The database is freely accessible and has been utilized in several publications.  We illustrate the use of this database and highlight the effects of CO upper limits on the inferred slopes of the local scaling relations between stellar mass, star formation rate (SFR), and H$_2$ surface densities.  We find that the correlation between H$_2$ and SFR surface density is the tightest among the three relations.

\end{abstract}


\section{Introduction} \label{sec:intro}

Over the past two decades, observational surveys have established a number of key scaling relations among galaxy properties. 
Of primary importance is the correlation between a galaxy's star formation rate (SFR) and its stellar mass ($M_*$), which for star-forming galaxies follows a tight relationship known as the star-forming ``main sequence'' \citep[SFMS;][]{Brinchmann:04,Salim:07}. 
The SFR in turn is expected to be closely linked to the supply of star-forming gas, a quantity which is traditionally probed with radio observations of \HI\ and CO emission, tracing the atomic and molecular gas respectively.
At the same time, studies which combine SFR and gas measurements have been critical to understanding the cycling of gas in galaxies, and form the basis of another important scaling relation, the ``star formation law,'' linking the SFR with gas content \citep{Kennicutt:12}.
Studies of the integrated gas content of galaxies, e.g.\ 
xCOLD GASS \citep{Saintonge:17}, have established that the molecular gas mass $M_{\rm H_2}$ is the main determinant of a galaxy's offset in SFR relative to the main sequence \citep{Saintonge:22}.
Thus, for main-sequence galaxies, all three of the quantities $M_{\rm H_2}$, $M_*$, and SFR are tightly correlated.
Spatially resolved studies of main-sequence galaxies, although limited to smaller samples, have established that these correlations are also observed among the mass surface densities of H$_2$, stars, and SFR within galaxies \citep{Wong:02,Bigiel:08,Wong:13,CanoDiaz:16,Lin:19}.

The Extragalactic Database for Galaxy Evolution (EDGE) survey \citep{Bolatto:17} was the first galaxy survey to couple CO interferometry, from the Combined Array for Research in Millimeter-wave Astronomy (CARMA), with a large optical program of integral-field spectroscopy (IFS), the Calar Alto Legacy Integral Field Area Survey (CALIFA) \citep{Sanchez:16}. 
CALIFA provides a full suite of spatially-resolved spectroscopic indicators for abundances, ionization, extinction-corrected star formation rate, stellar population ages, star formation histories, and kinematics \citep{Sanchez:Rx} for a well-defined sample that is designed to be representative of the $z$=0 universe \citep{Sanchez:12}.
{EDGE provides the key molecular gas information for a sub-sample of infrared-selected CALIFA galaxies.  Here we briefly summarize some of the main scientific highlights of EDGE.}
Globally, and consistent with previous work \citep{Saintonge:11}, the EDGE-CALIFA galaxies show a positive correlation between the molecular gas depletion time (\tdep) and $M_*$ \citep{Bolatto:17}, implying that more massive galaxies are less efficient at turning their observed molecular gas into stars. 
On resolved ($\sim$1--2 kpc) scales, the depletion time is also longer in early-type and massive galaxies \citep{Bolatto:17,Colombo:18,Villanueva:21}, but shows no clear relationship with stellar mass surface density (\sigstar) or the Toomre $Q$ instability parameter for stars and gas, and indeed both increases and decreases in \tdep\ with radius are observed \citep{Utomo:17}. 
These results underscore the influence of both local and global galaxy properties for the star formation efficiency. 
{The EDGE-CALIFA data also enable detailed studies of CO kinematics, which have been used to validate common approaches to deriving the circular velocity from stellar kinematics \citep{Leung:18}, and demonstrate the influence of thick ionized gas disks on H$\alpha$ rotation curves \citep{Levy:18,Levy:19}.
More recently, the EDGE-CALIFA collaboration has obtained additional CO observations of CALIFA galaxies with the Atacama Pathfinder Experiment (APEX) and the Atacama Compact Array (ACA), in \citet{Colombo:20}, \citet{Sanchez:21}, \citet{GaraySolis:23}, and \citet{Villanueva:23}, but these data sets are not discussed in the present paper.}

The need to efficiently manage both spatially resolved and global quantities spurred the design and development of a custom database for use within the EDGE collaboration.
The key requirement was to provide, for each galaxy in the sample, a set of global values, 1-D profiles and 2-D images, in separate tables that were easily joined for analysis. 
Over the course of the EDGE project, a number of intermediate databases have been generated, the most important being the SQL database used in \citet{Bolatto:17}, \citet{Dey:19}, and \citet{Cao:23}.  While SQL is a powerful and well-established database query language, the modest size of the EDGE repository led us to ultimately adopt a simpler approach based on Astropy tables \citep{Astropy:13, Astropy:18, Astropy:22}.  A pure Python implementation simplifies both the generation of and interaction with the database.  
The Python-based EDGE-CALIFA database ({\tt edge\_pydb}) presented in this paper is the successor to the SQL database, retaining essentially the same information aside from the use of a square rather than hexagonal sampling grid and important differences in handling of the CALIFA data described in \S \ref{sec:compare}.  Preliminary versions of the Python-based database have been used to produce results presented in \citet{Sanchez:21}, \citet{Barrera:21}, \citet{Villanueva:21}, and \citet{Ellison:21}.
An important aspect of our approach is that the format of the tables is largely secondary; we have found the Enhanced CSV (ECSV) format suitable for the global and 1-D profiles and the 
Hierarchical Data Format version 5 (HDF5) format suitable for 2-D images, but any format that supports preservation of metadata (including FITS) would also suffice.  While HDF5 offers improvements relative to FITS in terms of input/output speed and compression, this is of secondary importance for our purposes compared to offering a clear distinction between the original image data and the derived data tables.

In this paper we provide a detailed description of the construction of the database (\S\ref{sec:construct}), discuss some scientific applications (\S\ref{sec:applications}), and provide a comparison with the previous-generation EDGE database (\S\ref{sec:compare}).  We summarize our results in \S\ref{sec:summary}.

\section{Database Construction}\label{sec:construct}

\subsection{Input data products}\label{sec:inputdata}

\citet{Bolatto:17} presented the original EDGE-CALIFA sample (hereafter simply the ``EDGE sample''), consisting of 126 galaxies from Data Release 3 (DR3) of the CALIFA IFU survey \citep{Sanchez:16} that were observed in CO and \ttco\ by CARMA.\@ 
To briefly summarize, the CALIFA main sample of 667 galaxies was randomly selected for observation from a 937-galaxy ``mother sample'' 
derived from diameter {($r$ band isophotal major axis between 45\arcsec\ and 80\arcsec)} and redshift {($0.005 < z < 0.03$)} cuts applied to the seventh data release (DR7; \citealt{SDSS7}) 
of the Sloan Digital Sky Survey (SDSS).  The EDGE sub-sample was additionally selected to have high infrared flux (as measured by the {\sl WISE} 22$\mu$m survey, {\citealt{Wright:10}}) 
and a possible CO detection in CARMA's E configuration {(used to initially survey 177 galaxies)} was a prerequisite for continuing with CARMA observations in the D configuration.  As a result, the EDGE galaxies tend to have higher gas content and star formation activity compared to the overall CALIFA main sample.  More details about the EDGE sample and the CARMA observations are provided in \citet{Bolatto:17}.

For each galaxy we obtain three data cubes, an optical cube covering 4200--7100 \AA\ (unvignetted) {with a channel width of 2 \AA\ and a FWHM} resolution of 6 \AA\ (CALIFA `V500' setup), and CO and \ttco\ cubes covering a typical velocity range of 860 \kms\ (centered on the galaxy's systemic velocity) {with a channel spacing of 20 \kms\ derived from downsampling the original 3.4 \kms\ (CO) and 14.3 (\ttco)} \kms\ spectral channels. The field of view for CARMA EDGE extends to offsets of $\sim$50\arcsec\ (50\% sensitivity contour) while the hexagonal field of view for CALIFA extends to $\sim$35\arcsec.  CALIFA data for one galaxy, NGC 2486, were excluded from DR3 due to poor quality, and the corresponding CARMA data show only a marginal detection of CO.  We therefore drop this galaxy from the final database, leaving a sample of 125 galaxies. The \ttco\ cubes have much lower signal-to-noise (owing to the faintness of the line) and are analyzed in a separate paper \citep{Cao:23}.  While we focus on the optical and CO data in this paper, similar data products are available for \ttco.

We note that the native {spectral} resolution of the CO data is $\sim$5 \kms, and that the EDGE collaboration has also generated cubes with 10 \kms\ channel spacing.  These have been used for particular applications where velocity resolution is critical (e.g., studies of CO kinematics), but for most applications a 20 \kms\ channel spacing is preferred because of improved sensitivity per channel.  {The {\tt edge\_pydb} database is therefore based on cubes that employ a 20 \kms\ channel spacing.}  Similarly, higher spectral resolution CALIFA products are also available using the `V1200' spectral setup, but these are over a more limited velocity range (excluding the important H$\alpha$ line), so these are not included in the general-purpose database.

{While the CO cubes have a typical synthesized beam FWHM of 4\farcs5 \citep{Bolatto:17}, given the ellipticity of the beams we convolve to a common 7\arcsec\ curcular beam to produce the database.  At this resolution, the typical 3$\sigma$ sensitivity limit of the CO data is 3 K km/s, corresponding to 13 \Msol\ pc$^{-2}$ of molecular gas.  Matching to 7\arcsec\ resolution requires degrading the CALIFA images from their native resolution of approximately 2\farcs5 (FWHM), so we also construct a CALIFA-only database at the native resolution, as discussed further below.}

\subsection{Overall database architecture}

In constructing a database for legacy reuse, it became clear that the data can span a dimensionality from 0 (single value for a whole galaxy) to 3 (values that depend on sky position and radial velocity).  To minimize redundancy, it is best to store the data in its lowest dimensional form, and replicate it along additional axes only as needed.  Given our modest sample size, we found that tables of dimensionality 0 and 1 could effectively be stored as plain text, in the self-documenting Enhanced Character Separated Values (ECSV) format.  For the EDGE sample, a 0-dimensional table contains only $\sim$100 rows, and a 1-dimensional table (e.g. radial profiles, rotation curves, or integrated spectra) will increase the number of rows by a factor of 10--40, which is still manageable in textual format.  Higher dimensional tables are stored in a binary compressed format (HDF5).

Even for data of the same dimensionality, it can be advantageous to create separate data tables, corresponding to different data sets or analyses which may be updated over time.  For example, a table of global photometric parameters from SDSS can be maintained separately from similar parameters derived from CALIFA or the NASA Extragalactic Database (NED).  Alternatively, one may wish to create separate versions of a table corresponding to different resolutions of the same data, or different ways of masking a spectral data cube. Tables can then be joined as needed to avoid handling unnecessarily large tables.  To facilitating the joining process, we use standard names for the index columns that are used to match different tables, and we try to avoid duplication of column names for tables that are likely to be joined.

Finally, in addition to the tables themselves, we have created a Python {\tt EdgeTable} class which is a subclass of the {\tt astropy.table.Table} class.  This allows an experienced Python user to immediately get to work joining and plotting the tables.  We have added some basic plotting capabilities that are demonstrated in a set of accompanying Jupyter notebooks.  The class is made available through the Python package {\tt edge\_pydb}\footnote{\url{https://github.com/tonywong94/edge_pydb}} which can be installed via the standard Python Package Index (PyPI) package manager (using the {\tt pip} command).  The package includes the full Python source code needed to regenerate the database from FITS images.

We discuss the components of the database in further detail below.  We first describe the global parameter (0-dimensional) tables, followed by the 2- and 3-dimensional image tables, and finally the 1-dimensional profiles (which are generally based on analysis of the higher dimensional data).

\begin{figure}    
\includegraphics[width=\columnwidth]{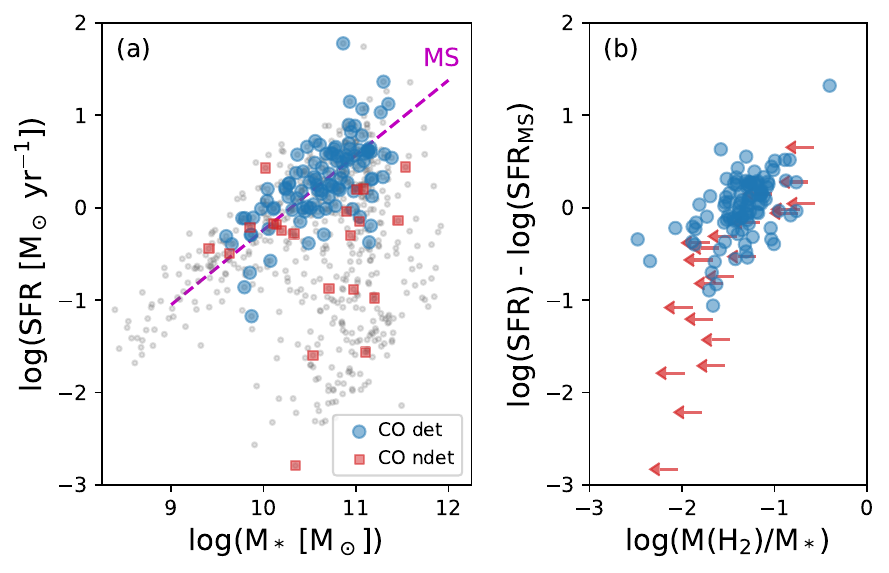}
\caption{(a) Location of EDGE-CALIFA sample in a plot of SFR vs.\ $M_*$.  The dashed magenta line shows the approximate locus of the star-forming main sequence from \citet{CanoDiaz:16}.  CO non-detections are indicated by red squares. {Small gray dots are used to represent the full CALIFA DR3 sample.} (b) Vertical offset from the star-forming main sequence as a function of molecular to stellar mass ratio.  The two quantities are correlated, underscoring the importance of gas content for continued star formation.  CO non-detections (2$\sigma$ upper limits) are indicated by horizontal arrows.}
\label{fig:sfms}
\end{figure}

\begin{deluxetable*}{ccc}
\tablecaption{Description of columns in the CALIFA global table.\label{tab:califa}}
\tablehead{\colhead{colname} & \colhead{units} & \colhead{description}}
\startdata
ID & & CALIFA ID number \\
Name & & Galaxy Name \\
caSu & $\mathrm{mag}$ & SDSS $u$ magnitude from CALIFA synthetic photometry\\
caSg & $\mathrm{mag}$ & SDSS $g$ magnitude from CALIFA synthetic photometry\\
caSr & $\mathrm{mag}$ & SDSS $r$ magnitude from CALIFA synthetic photometry\\
caSi & $\mathrm{mag}$ & SDSS $i$ magnitude from CALIFA synthetic photometry\\
caB & $\mathrm{mag}$ & $B$ magnitude from CALIFA synthetic photometry\\
caV & $\mathrm{mag}$ & $V$ magnitude from CALIFA synthetic photometry\\
caR & $\mathrm{mag}$ & $R$ magnitude from CALIFA synthetic photometry\\
caRe & arcsec & Equivalent galactocentric radius Re from CALIFA photometry\\
caeRe & arcsec & Error in equivalent galactocentric radius from CALIFA photometry\\
caEllipticity & & Ellipticity $\sqrt{(1-b^2/a^2)}$ from CALIFA photometry\\
caPA & degree & PA from CALIFA photometry\\
caR50 & arcsec & R50 from CALIFA photometry\\
caeR50 & arcsec & Error in R50 from CALIFA photometry\\
caR90 & arcsec & R90 from CALIFA photometry\\
caeR90 & arcsec & Error in R90 from CALIFA photometry\\
cazgas & & Redshift for gas lines from {\tt z\_gas} in get\_proc\_elines table\\
cazstars & & Redshift for stars from {\tt z\_stars} in get\_proc\_elines table\\
caAge & dex(Gyr) & Mean stellar age from {\tt log\_age\_mean\_LW} in get\_proc\_elines\\
caeAge & dex(Gyr) & Error in mean stellar age from {\tt s\_log\_age\_mean\_LW} in get\_proc\_elines\\
caFHa & dex($\mathrm{10^{-16}\,erg\,cm^{-2}\,s^{-1}}$) & Log of H$\alpha$ flux from {\tt log\_F\_Ha} in get\_proc\_elines\\
caFHacorr & dex($\mathrm{10^{-16}\,erg\,cm^{-2}\,s^{-1}}$) & Log of extinction corrected H$\alpha$ flux from {\tt log\_F\_Ha\_cor} in get\_proc\_elines\\
caLHacorr & dex($\mathrm{erg\,s^{-1}}$) & Log of extinction corrected H$\alpha$ luminosity from {\tt log\_L\_Ha\_cor} in get\_proc\_elines\\
caMstars & dex($\mathrm{M_{\odot}}$) & Log of stellar mass from {\tt log\_Mass} in get\_proc\_elines \\
caeMstars & dex($\mathrm{M_{\odot}}$) & Error in log of stellar mass from {\tt error\_Mass} in get\_proc\_elines\\
caSFR & dex($\mathrm{M_{\odot}\,yr^{-1}}$) & SFR from {\tt lSFR} in get\_proc\_elines\\
caeSFR & dex($\mathrm{M_{\odot}\,yr^{-1}}$) & Error in SFR from {\tt e\_lSFR} in get\_proc\_elines\\
caOH & $\mathrm{dex}$ & Oxygen abundance as 12+log(O/H) from {\tt OH\_O3N2} in get\_proc\_elines\\
caeOH & $\mathrm{dex}$ & Error in oxygen abundance from {\tt e\_OH\_O3N2} in get\_proc\_elines\\
caAvgas & $\mathrm{mag}$ & Nebular extinction as Av from {\tt Av\_gas\_LW\_Re} in get\_proc\_elines\\
caeAvgas & $\mathrm{mag}$ & Error in nebular extinction from {\tt e\_Av\_gas\_LW\_Re} in get\_proc\_elines\\
caAvstars & $\mathrm{mag}$ & Stellar extinction as Av from {\tt Av\_ssp\_stats\_mean} in get\_proc\_elines\\
caeAvstars & $\mathrm{mag}$ & Error in stellar extinction {\tt Av\_ssp\_stats\_stddev} in get\_proc\_elines\\
caDistP3d & $\mathrm{Mpc}$ & Luminosity distance in Mpc used by Pipe3D, from {\tt DL} in get\_proc\_elines\\
caDistMpc & $\mathrm{Mpc}$ & Luminosity distance in Mpc from {\tt cazgas} assuming $H_0$=70, $\Omega_m$=0.27, $\Omega_\Lambda$=0.73\\
caFlgWav5 & $\mathrm{}$ & Flag (-1/0/1/2=NA/good/minor/bad) for wavelength calibration in V500\\
caFlgWav12 & $\mathrm{}$ & Flag (-1/0/1/2=NA/good/minor/bad) for wavelength calibration in V1200\\
caFlgReg5 & $\mathrm{}$ & Flag (-1/0/1/2=NA/good/minor/bad) for 2D registration rel to SDSS in V500\\
caFlgReg12 & $\mathrm{}$ & Flag (-1/0/1/2=NA/good/minor/bad) for 2D registration rel to SDSS in V1200\\
caFlgImg5 & $\mathrm{}$ & Flag (-1/0/1/2=NA/good/minor/bad) for reconstructed image quality in V500\\
caFlgImg12 & $\mathrm{}$ & Flag (-1/0/1/2=NA/good/minor/bad) for reconstructed image quality in V1200%
\enddata
\end{deluxetable*}

\subsection{Global parameter tables}\label{sec:globpar}

The global parameter tables are organized into four groups, as described below.

\paragraph{{Characteristics of radio observations.}}  
These include the characteristics of the CO data such as beam size, integration time, channel noise, systemic velocity adopted for the observation, and integrated line fluxes for \HI, CO and \ttco\ emission. \HI\ fluxes in units of Jy \kms\ are derived from spectra obtained in GBT program 15B-287 (PI: D. Utomo) and from archival spectra published by \citet{vanDriel:01}, \citet{Springob:05}, and \citet{Masters:14}.  Details about the GBT observations will be presented in Wen et al.\ (2024, in preparation).  CO and \ttco\ fluxes are provided in the same units and are measured from cubes at both the native resolution ({\tt edge\_coflux\_natv.csv}) and a common resolution of 7\arcsec\ ({\tt edge\_coflux\_smo7.csv}).  Multiple CO tables with similar file names {(e.g., {\tt edge\_coobs\_D.csv}, {\tt edge\_coobs\_E.csv}, and {\tt edge\_coobs\_DE.csv})} are present because CARMA observations were conducted in the E configuration for 177 galaxies and in the D configuration for 126 galaxies, yielding a set of low-resolution cubes derived from E array data only, in addition to the principal set containing both D and E array data.  Furthermore, the D+E cubes were generated at two velocity resolutions (10 \kms\ and 20 \kms) and at two angular resolutions (native elliptical beam and a common 7\arcsec\ circular beam).  While characteristics of these different cubes are provided in separate ECSV files or table columns, the user should note that only the combined (D+E) products with a 7\arcsec\ beam and 20 \kms\ channels are provided as 2-D and 3-D image tables.

\paragraph{External databases.}  
These tables collect galaxy properties summarized in the LEDA\footnote{\url{http://leda.univ-lyon1.fr}} \citep{Makarov:14} and NED\footnote{\url{http://ned.ipac.caltech.edu}} databases, including coordinates, morphology, orientation parameters, diameter, redshift, and distance.  
Tables of {\sl WISE} photometry from \citet{Bitsakis:19}, corrected following the discussion in \citet[][their \S5.2]{Levy:19}, and Galactic extinction from \citet{Schlegel:98} are also provided.

The LEDA table is of particular importance because it provides the primary source for the coordinates of the galaxy centers, as well as the major axis position angles and inclinations, the latter derived from the apparent axis ratio using Equations (1) and (2) from \citet{Bottinelli:83}.  These orientation parameters are used to determine the galactocentric polar coordinates that are provided in the 2-D table (e.g., Table~\ref{tab:comom}).  A Python function ({\tt edge\_pydb.conversion.gc\_polr}) is provided to recalculate these coordinates based on different assumed orientation parameters.

\paragraph{CALIFA parameters.}  
These include galaxy properties derived from analyses conducted by the CALIFA consortium.  Examples include global measures of extinction, star formation rate, and stellar mass, effective radii from SDSS photometry, age and metallicity obtained from the {\tt Pipe3D} analysis pipeline \citep{Sanchez:Rx}, distances obtained from redshift, and CALIFA data quality flags from the Third Data Release \citep{Sanchez:16}.  {The contents of this table are summarized in Table~\ref{tab:califa}.  As noted in \citet{Sanchez:Rx}, {\tt Pipe3D} uses the {\tt gsd156} stellar library from \citet{CidFer:13}, adopting a \citet{Salpeter:55} stellar initial mass function (IMF), which we adopt for consistency throughout the database.}

Figure~\ref{fig:sfms}, derived from the CALIFA global table, shows the location of our sample galaxies in the conventional ``main sequence'' diagram of SFR vs.\ stellar mass ($M_*$), and the strong observed correlation between a galaxy's offset from the main sequence and the observed molecular gas to stellar mass ratio.  This figure demonstrates that molecular gas content has a strong effect on the location of a galaxy relative to the main sequence.

\paragraph{Derived parameters.}  
These include galaxy properties derived from analyses conducted by the EDGE team.  Examples include orientation parameters and systemic velocities from rotation curve fitting \citep{Leung:18, Levy:18} and half-light radii derived from integrating flux in elliptical rings \citep{Bolatto:17}.

\begin{figure*}[t!]
\includegraphics[width=\textwidth]{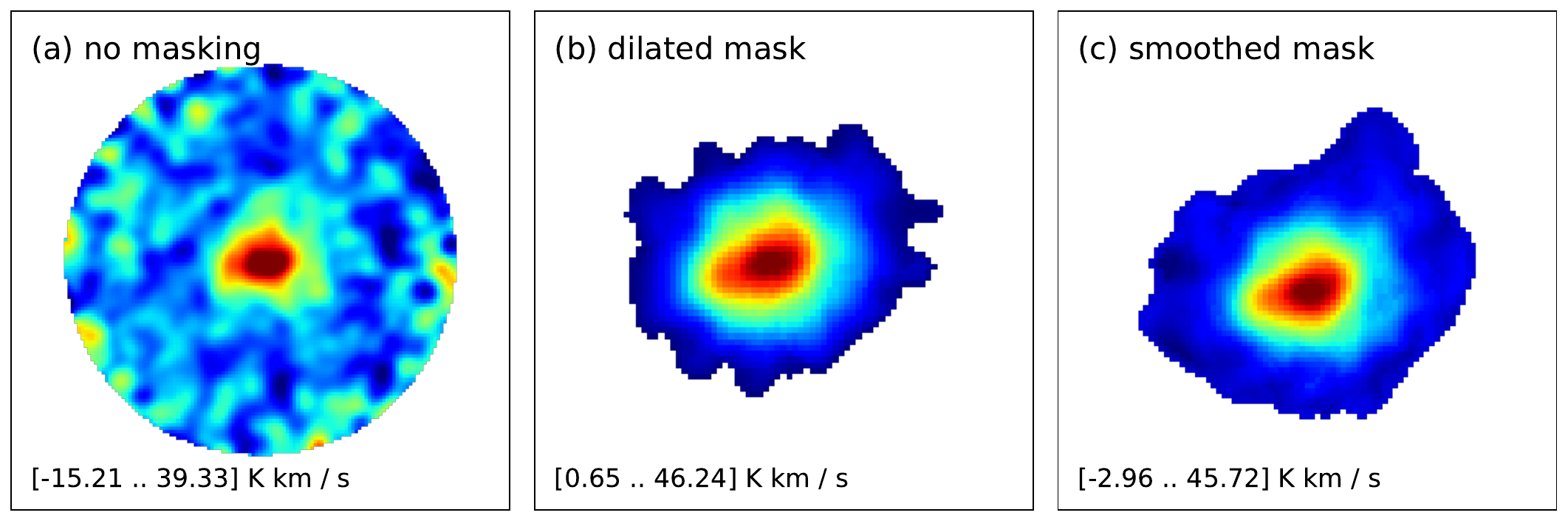}
\caption{Integrated CO intensity (moment-0) maps of NGC 4047, obtained with (a) no masking; (b) a dilated mask expanding from a 3.5$\sigma$ threshold over 2 consecutive channels out to a 2$\sigma$ threshold, and (c) a smoothed mask where the cube is convolved to 14\arcsec\ FWHM before a dilated mask is generated.  Each panel is approximately 147\arcsec\ square. The colormap range is shown at the bottom of each panel.}
\label{fig:mommaps}
\vspace*{1.5cm}
\end{figure*}

\subsection{CO image tables}\label{sec:cotable}

The image tables (saved in HDF5 format) are obtained by sampling FITS images.  For matching of the CARMA and CALIFA data, the coordinate grid is established by the original CO cubes, which were imaged with 1\arcsec\ pixels in an orthographic projection in J2000 coordinates.  The spatial extent of the cubes has been truncated to 128 $\times$ 128 pixels; this is sufficient to cover both the CARMA and CALIFA fields of view.  In addition to the CO cubes themselves, we sample moment images that are derived from the cubes following a masking process.  Three different masking approaches are employed, with the results from each approach saved in a separate table (`path') in the HDF5 file.  The paths are labeled as follows:
\begin{itemize}
    \item {\tt comom\_str}: This is a ``straight'' moment calculation (no masking employed).  Only the 0th moment (integrated intensity) is included, as the higher moments are largely unusable due to noise in the spectrum.  Even the 0th moment is typically dominated by noise and is useful primarily as a rough guide to what signal would be clearly detected in the absence of any masking. 
    \item {\tt comom\_dil}: Data are first masked using a ``dilated'' approach (expanding from a 3.5$\sigma$ threshold over 2 consecutive channels out to a 2$\sigma$ threshold, with a minimum area of 2 synthesized beams for each mask region) before moments are calculated.  For \ttco\ the dilated mask from the CO cube is applied.  This table also includes peak signal-to-noise ratio images as {\tt snrpk\_12} and  {\tt snrpk\_13}.  Although these are constructed without masking (and should thus logically be included in the {\tt comom\_str} table), they are included in this table as most users will adopt the dilated masks for initial data inspection.
    \item {\tt comom\_smo}: This is a ``pre-smoothing'' approach where the cube is first convolved to a resolution of 14\arcsec\ before generating a dilated mask as described above.  This generally yields a more permissive mask than  {\tt comom\_dil}, which is then applied to the unsmoothed data before the moments are calculated.  For \ttco\ the corresponding mask from the CO cube is applied.
\end{itemize}

For both ``dilated'' and ``smoothed'' approaches, an additional mask expansion of 1 pixel in all 3 dimensions (2 pixels for the smoothed approach) is introduced to pick up additional weak emission.  Since the masks are 3-D they are included with the cube data rather than the image tables.  The overall approach to moment map generation remains as described by \citet{Bolatto:17}, but using a Python implementation ({\tt maskmoment}\footnote{\url{https://github.com/tonywong94/maskmoment}}) to remove a previous dependency on IDL.
    
Figure~\ref{fig:mommaps} shows how the masking methods compare for NGC 4047.  The tables above also include estimates of molecular gas surface density ({\tt sigmol}) derived from the CO moment-0 values by applying a constant CO-to-H$_2$ conversion factor (see \S\ref{sec:sigmol}).  Note that tabulated intensities and surface densities have {\it not} been deprojected to face-on values, although a {\tt cosi} column (representing the cosine of the inclination) is provided which can be multiplied by any column to yield a deprojected value based on the LEDA derived inclination (\S\ref{sec:globpar}).  A summary of the columns currently provided in the {\tt comom\_dil} table is provided in Table~\ref{tab:comom}.

\subsection{CALIFA image tables}

For the CALIFA data, we sample a subset of the Pipe3D analysis products rather than the IFU cubes directly.  Separate tables (``paths'') for the Pipe3D products labeled as {\tt ELINES}, {\tt flux\_elines}, {\tt SFH}, {\tt SSP}, and {\tt indices} (see \citealt{Sanchez:Rx}) are provided.
Each of these products is ingested as a FITS file (or as an extension in a multi-extension FITS file) which is essentially a concatenation of individual images into a pseudo-cube.
Tables~\ref{tab:ssp} and \ref{tab:felines} summarize the columns that are provided in the  {\tt SSP} and {\tt flux\_elines} tables.
Pipe3D was run on two sets of cubes, the native-resolution DR3 cubes and a set of cubes obtained by convolving the DR3 cubes to a Gaussian PSF (7\arcsec\ FWHM) to match the CARMA data. To perform the convolution, we model the CALIFA PSF as a Moffat profile with the characteristics provided in the FITS header; in cases where this was not available, we assume a Moffat profile with FWHM 2\farcs5 and $\beta_{\rm M}$=2.15, which represent the mean and weighted median of the CALIFA DR3 sample as measured by \citet{Sanchez:16}. We store the native and convolved data in separate HDF5 files as detailed below.
The images that are matched to the CARMA resolution have been interpolated onto the CARMA astrometric grid, using nearest-neighbor sampling in the {\tt reproject} Python package, before pixel extraction.

\begin{deluxetable*}{cccc}
\tablecaption{Description of columns in the comom{\_}dil table.\label{tab:comom} }
\tablehead{\colhead{colname} & \colhead{units} & \colhead{description} & \colhead{n{\_}bad\tablenotemark{a}{}}}
\startdata
Name & & Galaxy Name & 0 \\
ix & & 0-based pixel index in x direction & 0 \\
iy & & 0-based pixel index in y direction & 0 \\
ra{\_}abs & degree & sample R.A. coordinate & 0 \\
dec{\_}abs & degree & sample Decl. coordinate & 0 \\
ra{\_}off & degree & R.A. offset from reference pixel & 0 \\
dec{\_}off & degree & Decl. offset from reference pixel & 0 \\
rad{\_}arc & arcsec & galactocentric radius based on LEDA & 0 \\
azi{\_}ang & degree & azimuthal angle based on LEDA & 0 \\
snrpk{\_}12 & & CO peak signal to noise ratio & 705 \\
mom0{\_}12 & $\mathrm{K\,km\,s^{-1}}$ & CO integrated intensity using dilated mask & 1657 \\
e{\_}mom0{\_}12 & $\mathrm{K\,km\,s^{-1}}$ & CO error in mom0 assuming dilated mask & 1657 \\
mom1{\_}12 & $\mathrm{km\,s^{-1}}$ & CO intensity weighted mean velocity using dilated mask & 1670 \\
e{\_}mom1{\_}12 & $\mathrm{km\,s^{-1}}$ & CO error in mom1 assuming dilated mask & 1670 \\
mom2{\_}12 & $\mathrm{km\,s^{-1}}$ & CO intensity weighted vel disp using dilated mask & 1694 \\
e{\_}mom2{\_}12 & $\mathrm{km\,s^{-1}}$ & CO error in mom2 assuming dilated mask & 1694 \\
sigmol & $\mathrm{M_{\odot}\,pc^{-2}}$ & apparent H$_2$+He surf density not deprojected & 1657 \\
e{\_}sigmol & $\mathrm{M_{\odot}\,pc^{-2}}$ & error in sigmol not deprojected & 1657 \\
cosi & $\mathrm{}$ & factor to deproject to face-on using ledaAxIncl & 0 \\
snrpk{\_}13 & $\mathrm{}$ & $^{13}$CO peak signal to noise ratio & 648 \\
mom0{\_}13 & $\mathrm{K\,km\,s^{-1}}$ & $^{13}$CO integrated intensity using dilated mask & 1657 \\
e{\_}mom0{\_}13 & $\mathrm{K\,km\,s^{-1}}$ & $^{13}$CO error in mom0 assuming dilated mask & 1657 \\
mom1{\_}13 & $\mathrm{km\,s^{-1}}$ & $^{13}$CO intensity weighted mean velocity using dilated mask & 1673 \\
e{\_}mom1{\_}13 & $\mathrm{km\,s^{-1}}$ & $^{13}$CO error in mom1 assuming dilated mask & 1670 \\
mom2{\_}13 & $\mathrm{km\,s^{-1}}$ & $^{13}$CO intensity weighted vel disp using dilated mask & 1731 \\
e{\_}mom2{\_}13 & $\mathrm{km\,s^{-1}}$ & $^{13}$CO error in mom2 assuming dilated mask & 1731%
\enddata
\tablenotetext{a}{Number of undefined pixels for NGC 4047, out of a maximum of $43^2=1849$.  This will differ among galaxies and is only listed for illustrative purposes.}
\end{deluxetable*}%

\begin{deluxetable*}{cccc}
\tablecaption{Description of columns in the SSP table.\label{tab:ssp}}
\tablehead{\colhead{colname} & \colhead{units} & \colhead{description} & \colhead{n{\_}bad}}
\startdata
Name & $\mathrm{}$ & Galaxy Name & 0 \\
ix & $\mathrm{}$ & 0-based pixel index in x direction & 0 \\
iy & $\mathrm{}$ & 0-based pixel index in y direction & 0 \\
{\dots} & {\dots} & {\dots} & {\dots} \\
Vcont{\_}ssp{\_}sm & $\mathrm{10^{-16}\,erg\,\mathring{A}^{-1}\,s^{-1}\,cm^{-2}}$ & pseudo V-band map & 1395 \\
cont{\_}segm{\_}sm & $\mathrm{}$ & continuum segmentation file & 1395 \\
cont{\_}dezon{\_}sm & $\mathrm{}$ & continuum dezonification file & 1395 \\
medflx{\_}ssp{\_}sm & $\mathrm{10^{-16}\,erg\,\mathring{A}^{-1}\,s^{-1}\,cm^{-2}}$ & median intensity flux within the wavelength range & 1395 \\
e{\_}medflx{\_}ssp{\_}sm & $\mathrm{10^{-16}\,erg\,\mathring{A}^{-1}\,s^{-1}\,cm^{-2}}$ & error in median intensity flux within the wavelength range & 1395 \\
age{\_}lwt{\_}sm & dex(yr) & luminosity weighted age of the stellar population & 1395 \\
age{\_}mwt{\_}sm & dex(yr) & mass weighted age of the stellar population & 1395 \\
e{\_}age{\_}lwt{\_}sm & $\mathrm{}$ & error in age of the stellar population & 1395 \\
ZH{\_}lwt{\_}sm & $\mathrm{dex}$ & luminosity weighted metallicity of the stellar population & 1395 \\
ZH{\_}mwt{\_}sm & $\mathrm{dex}$ & mass weighted metallicity of the stellar population & 1395 \\
e{\_}ZH{\_}lwt{\_}sm & $\mathrm{}$ & error in metallicity of the stellar population & 1395 \\
Av{\_}ssp{\_}sm & $\mathrm{mag}$ & average dust attenuation of the stellar population & 1395 \\
e{\_}Av{\_}ssp{\_}sm & $\mathrm{mag}$ & error in the average dust attenuation of the stellar population & 1395 \\
vel{\_}ssp{\_}sm & $\mathrm{km\,s^{-1}}$ & velocity of the stellar population & 1395 \\
e{\_}vel{\_}ssp{\_}sm & $\mathrm{km\,s^{-1}}$ & error in the velocity of the stellar population & 1395 \\
vdisp{\_}ssp{\_}sm & $\mathrm{km\,s^{-1}}$ & velocity dispersion of the stellar population & 1395 \\
e{\_}vdisp{\_}ssp{\_}sm & $\mathrm{km\,s^{-1}}$ & error in velocity dispersion of the stellar population & 1395 \\
ML{\_}ssp{\_}sm & $\mathrm{M_{\odot}\,L_{\odot}^{-1}}$ & average mass-to-light ratio of the stellar population & 1395 \\
mass{\_}ssp{\_}sm & dex($\mathrm{M_{\odot}\,arcsec^{-2}}$) & stellar mass density & 1395 \\
mass{\_}Avcor{\_}ssp{\_}sm & dex($\mathrm{M_{\odot}\,arcsec^{-2}}$) & stellar mass density dust corrected & 1395 \\
sigstar{\_}sm & $\mathrm{M_{\odot}\,pc^{-2}}$ & stellar mass surface density & 1395 \\
sigstar{\_}Avcor{\_}sm & $\mathrm{M_{\odot}\,pc^{-2}}$ & stellar mass surface density dust corrected & 1395 \\
fe{\_}medflx{\_}sm & $\mathrm{}$ & fractional error in continuum flux & 1395 \\
sigsfr{\_}ssp{\_}sm & $\mathrm{M_{\odot}\,Gyr^{-1}\,pc^{-2}}$ & Sigma{\_}SFR from $<$ 32 Myr SSP & 1395 \\
sigsfr{\_}Avcor{\_}ssp{\_}sm & $\mathrm{M_{\odot}\,Gyr^{-1}\,pc^{-2}}$ & Sigma{\_}SFR Av-corrected from $<$ 32 Myr SSP & 1395%
\enddata
\end{deluxetable*}%

\begin{deluxetable*}{cccc}
\tablecaption{Description of columns in the flux{\_}elines table.\label{tab:felines}}
\tablehead{\colhead{colname} & \colhead{units} & \colhead{description} & \colhead{n{\_}bad}}
\startdata
Name & $\mathrm{}$ & Galaxy Name & 0 \\
ix & $\mathrm{}$ & 0-based pixel index in x direction & 0 \\
iy & $\mathrm{}$ & 0-based pixel index in y direction & 0 \\
{\dots} & {\dots} & {\dots} & {\dots} \\
flux{\_}Hbeta{\_}sm & $\mathrm{10^{-16}\,erg\,s^{-1}\,cm^{-2}}$ & H$\beta$ intensity & 1386 \\
flux{\_}Halpha{\_}sm & $\mathrm{10^{-16}\,erg\,s^{-1}\,cm^{-2}}$ & H$\alpha$ intensity & 1386 \\
vel{\_}Hbeta{\_}sm & $\mathrm{km\,s^{-1}}$ & H$\beta$ velocity & 1386 \\
vel{\_}Halpha{\_}sm & $\mathrm{km\,s^{-1}}$ & H$\alpha$ velocity & 1386 \\
disp{\_}Hbeta{\_}sm & $\mathrm{\mathring{A}}$ & H$\beta$ velocity dispersion & 1386 \\
disp{\_}Halpha{\_}sm & $\mathrm{\mathring{A}}$ & H$\alpha$ velocity dispersion & 1386 \\
EW{\_}Hbeta{\_}sm & $\mathrm{\mathring{A}}$ & H$\beta$ equivalent width & 1386 \\
EW{\_}Halpha{\_}sm & $\mathrm{\mathring{A}}$ & H$\alpha$ equivalent width & 1386 \\
e{\_}flux{\_}Hbeta{\_}sm & $\mathrm{10^{-16}\,erg\,s^{-1}\,cm^{-2}}$ & error in H$\beta$ intensity & 1386 \\
e{\_}flux{\_}Halpha{\_}sm & $\mathrm{10^{-16}\,erg\,s^{-1}\,cm^{-2}}$ & error in H$\alpha$ intensity & 1386 \\
e{\_}vel{\_}Hbeta{\_}sm & $\mathrm{km\,s^{-1}}$ & error in H$\beta$ velocity & 1386 \\
e{\_}vel{\_}Halpha{\_}sm & $\mathrm{km\,s^{-1}}$ & error in H$\alpha$ velocity & 1386 \\
e{\_}disp{\_}Hbeta{\_}sm & $\mathrm{\mathring{A}}$ & error in H$\beta$ velocity dispersion & 1386 \\
e{\_}disp{\_}Halpha{\_}sm & $\mathrm{\mathring{A}}$ & error in H$\alpha$ velocity dispersion & 1386 \\
e{\_}EW{\_}Hbeta{\_}sm & $\mathrm{\mathring{A}}$ & error in H$\beta$ equivalent width & 1386 \\
e{\_}EW{\_}Halpha{\_}sm & $\mathrm{\mathring{A}}$ & error in H$\alpha$ equivalent width & 1386 \\
flux{\_}Hbeta{\_}sm3{\_}sm & $\mathrm{10^{-16}\,erg\,s^{-1}\,cm^{-2}}$ & H$\beta$ intensity after 3 pix smooth & 1386 \\
flux{\_}Halpha{\_}sm3{\_}sm & $\mathrm{10^{-16}\,erg\,s^{-1}\,cm^{-2}}$ & H$\alpha$ intensity after 3 pix smooth & 1386 \\
flux{\_}sigsfr0{\_}sm & $\mathrm{M_{\odot}\,Gyr^{-1}\,pc^{-2}}$ & SFR surface density no extinction & 1386 \\
e{\_}flux{\_}sigsfr0{\_}sm & $\mathrm{M_{\odot}\,Gyr^{-1}\,pc^{-2}}$ & error of uncorrected SFR surface density & 1386 \\
flux{\_}sigsfr{\_}corr{\_}sm & $\mathrm{M_{\odot}\,Gyr^{-1}\,pc^{-2}}$ & BD corrected SFR surface density & 1387 \\
e{\_}flux{\_}sigsfr{\_}corr{\_}sm & $\mathrm{M_{\odot}\,Gyr^{-1}\,pc^{-2}}$ & error of BD corrected SFR surface density & 1387 \\
flux{\_}AHa{\_}corr{\_}sm & $\mathrm{mag}$ & H$\alpha$ extinction from BD & 1387 \\
e{\_}flux{\_}AHa{\_}corr{\_}sm & $\mathrm{mag}$ & error of H$\alpha$ extinction from BD & 1387 \\
flux{\_}AHa{\_}smooth3{\_}sm & $\mathrm{mag}$ & H$\alpha$ extinction after 3 pix smooth & 1386 \\
flux{\_}sigsfr{\_}adopt{\_}sm & $\mathrm{M_{\odot}\,Gyr^{-1}\,pc^{-2}}$ & smooth+clip BD corrected SFR surface density & 1386 \\
BPT{\_}sm & $\mathrm{}$ & BPT type (-1=SF 0=inter 1=LINER 2=Sy) & 1397 \\
p{\_}BPT{\_}sm & $\mathrm{}$ & BPT probability & 1397 \\
SF{\_}BPT{\_}sm & $\mathrm{}$ & True if star forming (BPT=-1 and EW{\_}Ha$>$6) & 0 \\
ZOH{\_}sm & $\mathrm{dex}$ & 12+log(O/H) using O3N2 method in Marino+13 & 1489 \\
e{\_}ZOH{\_}sm & $\mathrm{dex}$ & error in 12+log(O/H) using O3N2 method in Marino+13 & 1489%
\enddata
\end{deluxetable*}%

\subsubsection{Full pixel databases}

For users that wish to recover all pixels from the FITS images, at the cost of significant oversampling, we provide large HDF5 files that contain the complete set of 1 arcsec$^2$ pixel values in tabular format.  
The {\tt edge\_carma\_allpix.pipe3d.hdf5} file contains all pixels in the native-resolution Pipe3D images, without regridding to match the CARMA astrometry.
The {\tt edge\_carma\_allpix.2d\_smo7.hdf5} file, which was used to generate Figure~\ref{fig:mommaps}, contains all pixels in the matched (7\arcsec\ FWHM) resolution CO and Pipe3D images, with CARMA and CALIFA data placed in separate ``paths'' (tables) in the HDF5 file.

\subsubsection{Square grid downsampling}

Given that a Gaussian beam with FWHM 7\arcsec\ spans an area of 55.5 arcsec$^2$, we are heavily oversampling the beam with our 1\arcsec\ pixels.  We therefore generate smaller tables which sample one of every three pixels in R.A. and Dec.\ (3\arcsec\ spacing), reducing the number of pixels per galaxy from $128^2$ to $43^2$ (a factor of 8.9).  We still cover each beam area with slightly more than 6 sampling pixels.  Combining the 125 galaxies, the image tables contain $43^2 \times 125 = 231\,125$ rows of sampled image data.  The {\tt edge\_carma.2d\_smo7.hdf5} file contains these extracted pixels from the matched-resolution CO and Pipe3D images. The {\tt edge\_carma.cocube\_smo7.hdf5} file contains the corresponding CO spectra and CO mask information at the extracted pixels.

\subsubsection{Hexagonal grid sampling}

\begin{figure}
    \includegraphics[width = 0.3 \paperwidth, keepaspectratio]{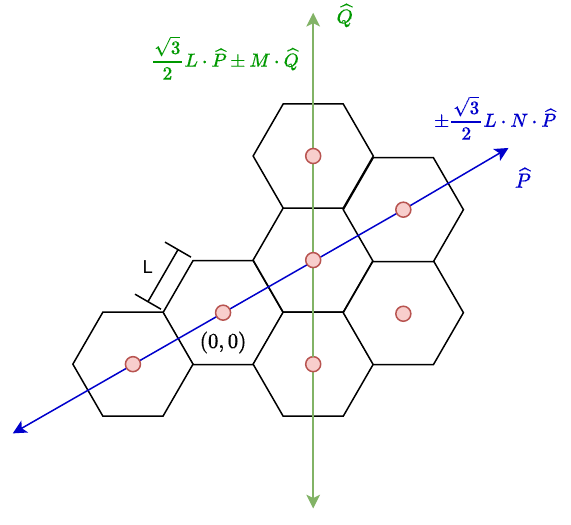}
    \centering
    \caption{The construction of the hexagonal grid. $L$ is the side length of the hexagon and can be set by the user. Blue and green vectors form the basis used to construct the grid. Hexagons are first are created along the diagonal, following the blue vector $\hat{P}$. Then, for each hexagon along $\hat{P}$, additional hexagons are created following the green vertical vector $\hat{Q}$.}
    \label{fig:hexgrid}
\end{figure}

\begin{figure*}
    \includegraphics[width=\textwidth]{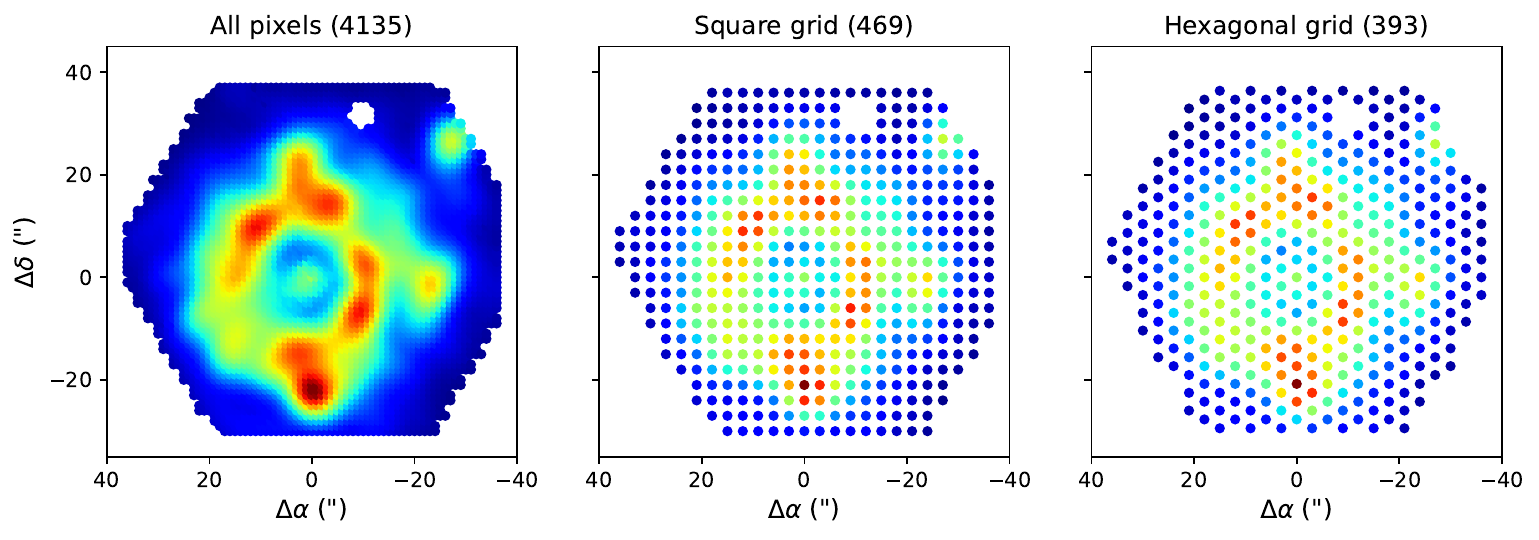}
    \caption{Comparison of the three sampling methods in the edge\_pydb, using the H$\alpha$ image of NGC 3687 as an example.  Counts of unblanked pixels are given in parentheses.}
    \label{fig:dotplot}
\end{figure*}

In addition to the square grid, we have also implemented an option to sample the {CARMA and CALIFA} data on a hexagonal grid (file {\tt edge\_carma\_hex.2d\_smo7.hdf5}).  For a uniform hexagonal grid, the minimum spacing between grid points can be increased by a factor of $2/\sqrt{3}$ compared to a square grid while still achieving the same level of oversampling.  Multiplying this factor by 3\arcsec\ yields a 3.5\arcsec\ spacing between adjacent hexagonal cells.  The reference pixel of the CO image is placed at the (0,0) position of the grid. The grid is constructed using two vectors as shown in Figure~\ref{fig:hexgrid}.  Hexagonal cells are created first along the diagonal line and then created in the vertical directions.

Interpolation of the original square pixel values onto the grid is done by calculating the distance from the center of the hexagon to the centers of the four nearest pixels, then weighting the pixel values by inverse distance.  This approach differs slightly from the hexagonally sampled SQL database, which also uses a 3\farcs5 nearest-neighbor spacing but does not apply interpolation, instead adopting the value of the 1\arcsec\ pixel closest to each hexagonal grid point.  A comparison of the three different sampling approaches (all pixels, square grid sampling, and hexagonal grid sampling) is provided in Figure~\ref{fig:dotplot}. {The comparison shows that the latter two sampling methods provide largely equivalent information to full-pixel sampling in a much more compact data format.}

\subsection{Derived quantities and uncertainties}

In addition to the CO intensity data and the Pipe3D output, we include additional derived columns in the database as described below.

\subsubsection{Star formation surface density}\label{sec:sfrdens}

The star formation surface density $\Sigma_{\rm SFR}$ is derived from H$\alpha$ surface brightness using the conversion factor from Equation 5 of \citet{Calzetti:10}, but with the numerical coefficient scaled by 1.51 to convert from a \citet{Kroupa:93} IMF to the \citet{Salpeter:55} IMF employed by Pipe3D: 
\begin{equation}
{\rm SFR} [M_{\odot}\;{\rm yr}^{-1}] = 8.23 \times 10^{-42}\ L({\rm H}\alpha)\;\rm [erg\; s^{-1}]
\end{equation}
Three separate estimates are calculated.  The first estimate ({\tt sigsfr0}) assumes no extinction and simply rescales the H$\alpha$ surface brightness, converting it to a luminosity surface density assuming each patch of the galaxy radiates isotropically.  The fractional error in this quantity is just the fractional error in the H$\alpha$ surface brightness.  The second estimate ({\tt sigsfr\_corr}) corrects the SFR for extinction, $A({\rm H\alpha})$, as obtained from the Balmer decrement \citep[e.g.,][]{Catalan:15}:
\begin{equation}
A({\rm H\alpha}) = \frac{K_{\rm H\alpha}}{-0.4(K_{\rm H\alpha}-K_{\rm H\beta})} \times \log_{10} \frac{I({\rm H\alpha})}{2.86\,I({\rm H\beta})},
\end{equation}
where we adopt $K_{\rm H\alpha} = 2.53$ and $K_{\rm H\beta} = 3.61$ based on the extinction curve of \citet{Cardelli:89}.  The uncertainty in this quantity is determined by propagating the errors of the H$\alpha$ and H$\beta$ intensities assuming these errors are uncorrelated. The third estimate ({\tt sigsfr\_adopt}) applies spatial smoothing (by a 2D Gaussian kernel with a standard deviation of 3\arcsec) to the H$\alpha$ and H$\beta$ images before their ratio is taken to obtain the Balmer decrement.  Smoothing reduces noise in the resulting estimate of $A({\rm H\alpha})$, increasing the fraction of inferred $A({\rm H\alpha})$ values between 0 and 2 mag from 54\% to 60\%; for spaxels classified as star-forming, this fraction increases from 78\% to 85\%.  This comes at the cost of some inconsistency between the coarser resolution $A({\rm H\alpha})$ images and the 7\arcsec\ resolution H$\alpha$ images that are being corrected.  We use the unsmoothed error maps for H$\alpha$ and H$\beta$ when propagating the uncertainty for {\tt sigsfr\_adopt}.  

None of the error estimates for $\Sigma_{\rm SFR}$ take into account uncertainties in the extinction curve, IMF, or other factors needed to convert from recombination line flux to star formation rate.  When the inferred 
$A({\rm H\alpha})<0$, we do not make an extinction correction, instead falling back on {\tt sigsfr0} and its uncertainty.  Values are blanked (set to NaN) when the inferred $A({\rm H\alpha})>6$.

\begin{figure*}
    \centering
    \includegraphics[width = 0.32 \textwidth]{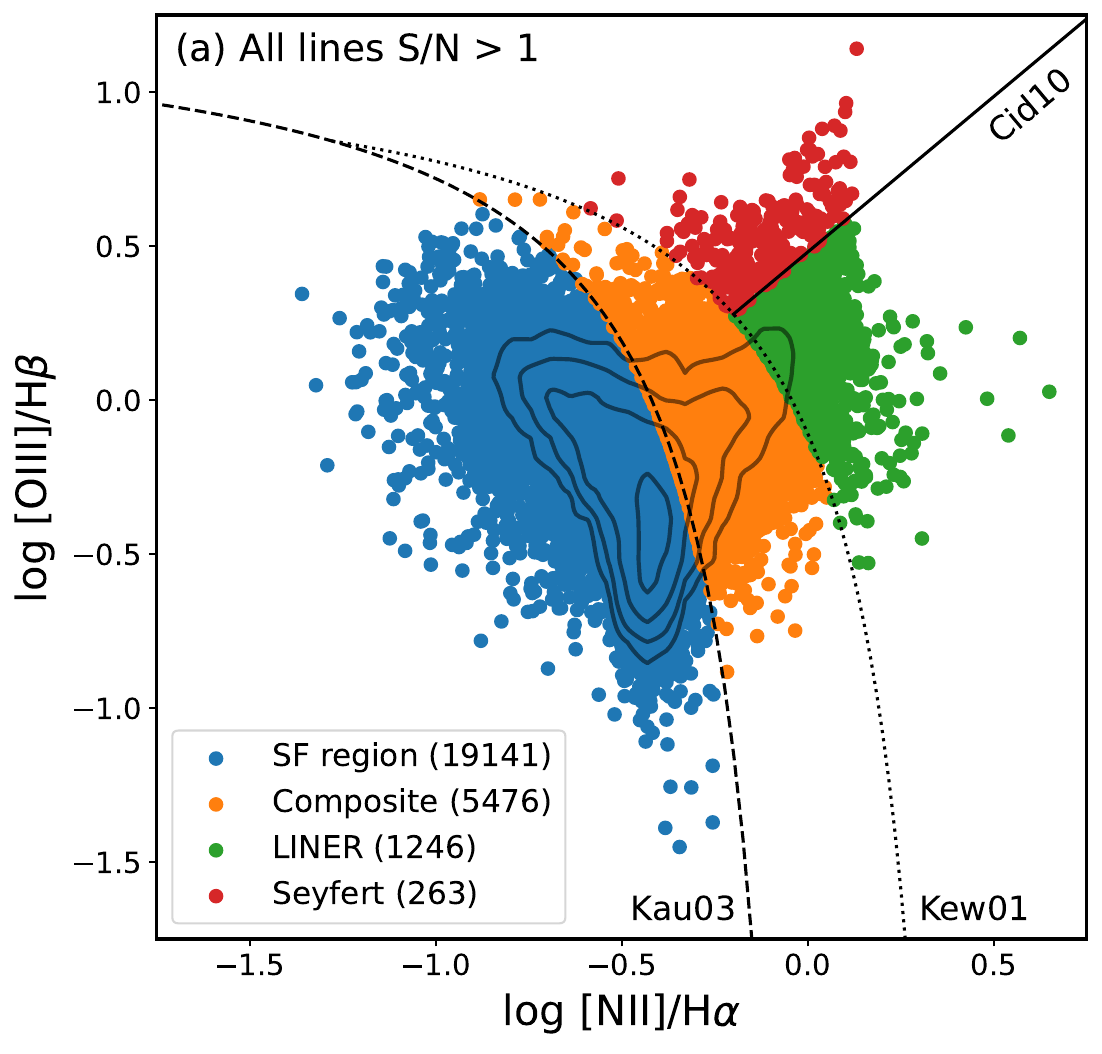}\hfill
    \includegraphics[width = 0.32 \textwidth]{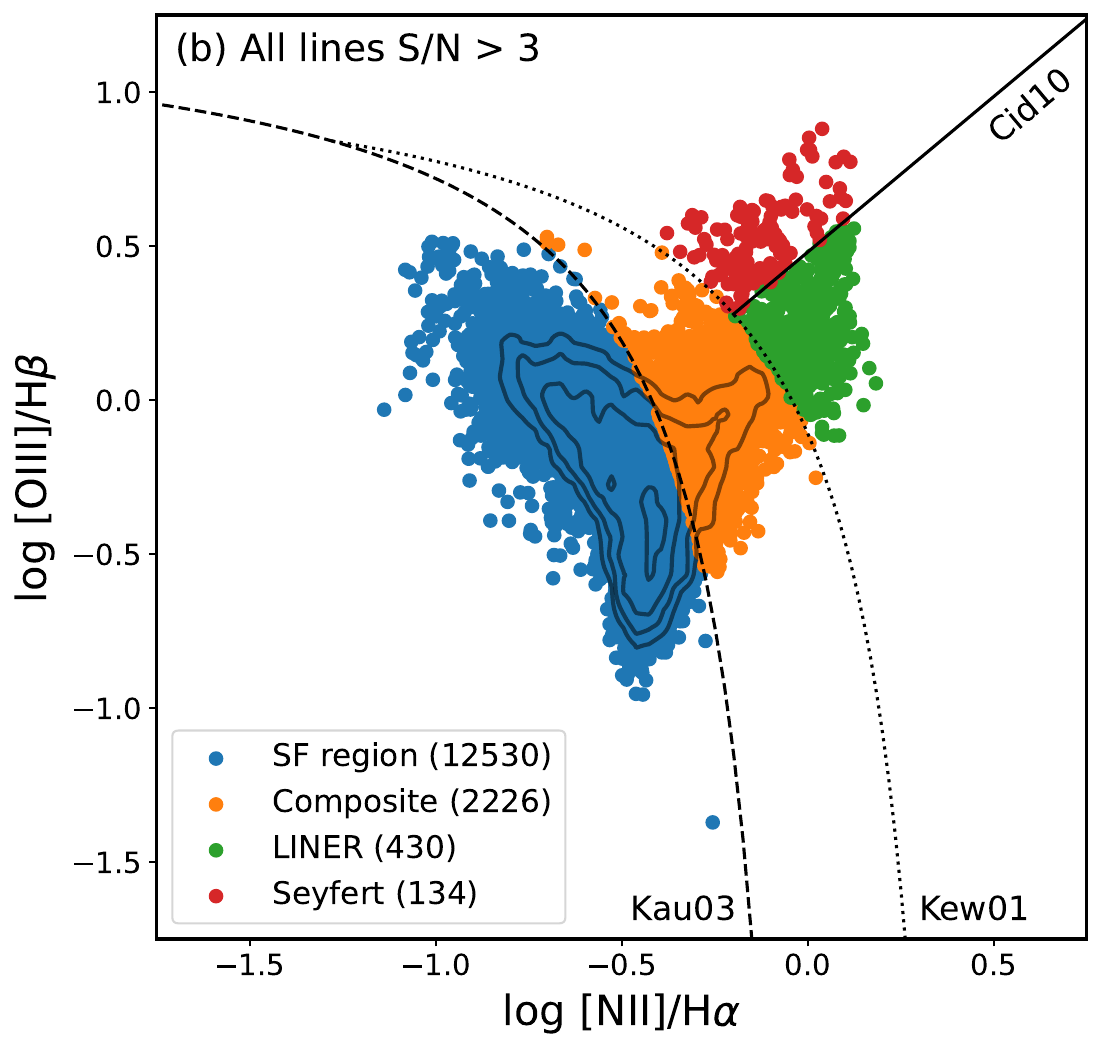}\hfill
    \includegraphics[width = 0.32 \textwidth]{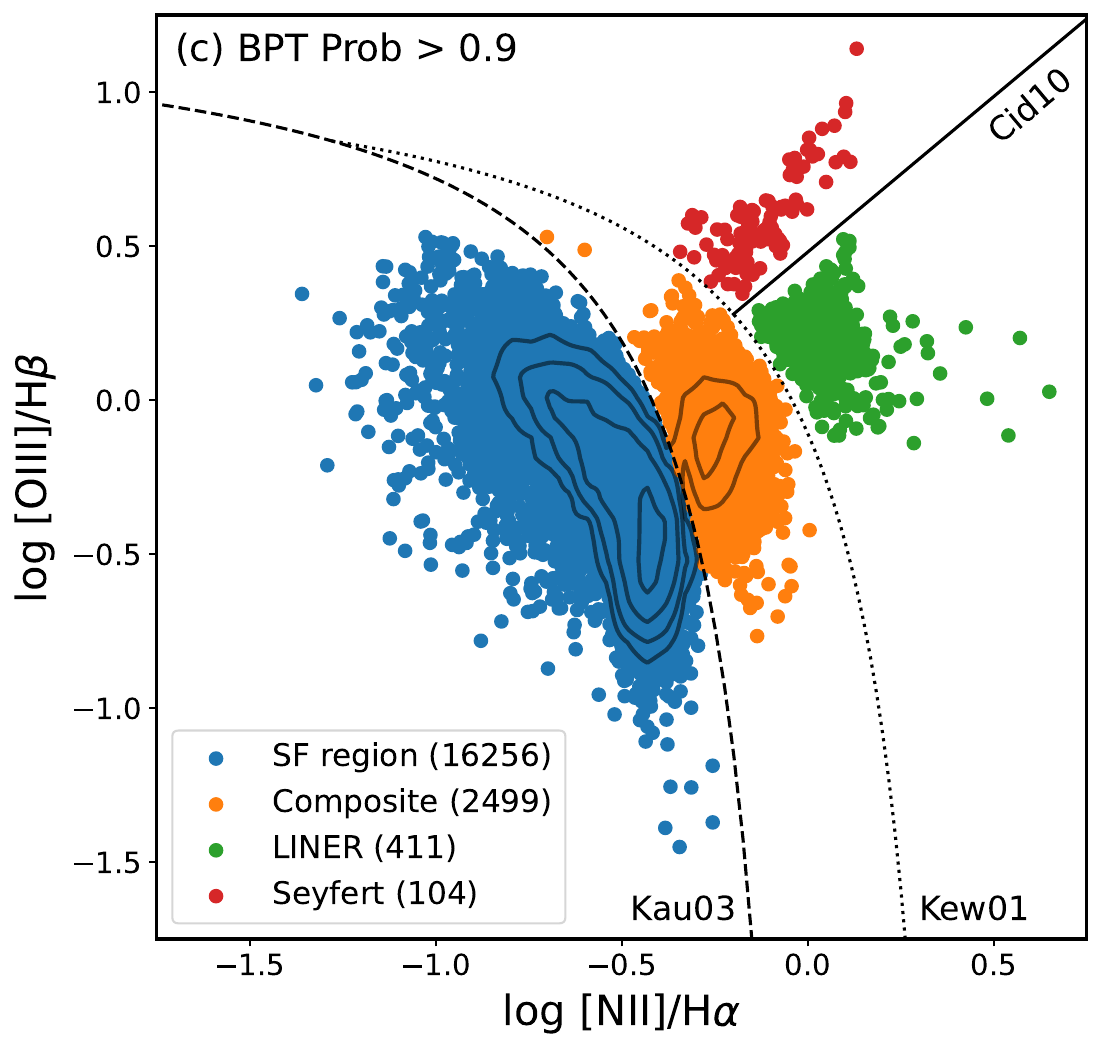}
    \caption{Comparison between BPT plots with different selection cuts applied. (a) All EDGE samples plotted where each of the four emission line fluxes exceeds its 1$\sigma$ uncertainty.  (b) Points from (a) where each line's flux exceeds 3$\sigma$.  (c) Points from (a) where the BPT probability estimate exceeds 0.9.  Gray contours indicate the density of points.  Demarcation lines defined by \citet{Kauffmann:03}, \citet{Kewley:01}, and \citet{CidFer:10} are shown as dashed, dotted, and solid curves respectively.}
    \label{fig:bpt_comp}
\end{figure*}

\subsubsection{Stellar surface density}

The {\tt SSP} table provided by Pipe3D provides a column {\tt mass\_ssp} that provides the stellar mass derived by Pipe3D for each 1 arcsec$^{2}$ CALIFA spaxel.  We add an additional {\tt sigstar} column that converts this value to $M_\odot$ pc$^{-2}$ using the galaxy distance adopted by Pipe3D.
Corresponding dust-corrected estimates {\tt mass\_AVcor\_ssp} and {\tt sigstar\_Avcor} are also provided; these make use of $A_V$ derived from the stellar population analysis, which can differ significantly from the Balmer decrement derived extinction.
Although Pipe3D does not provide formal uncertainties for {\tt mass\_ssp} or {\tt mass\_AVcor\_ssp}, the fractional error in the continuum flux ({\tt fe\_medflux}) can be used as a rough estimate.

The resolved star formation history provided by the {\tt SFH} table (in bins of age and metallicity) is also used to provide two estimates of $\Sigma_{\rm SFR}$ that are included in the {\tt SSP} table.
These estimates are computed by taking the mass fraction at ages $<$32 Myr (summed over metallicity), scaling by either {\tt sigstar} or {\tt sigstar\_Avcor}, and dividing the result by 32 Myr.  The choice of summing time bins up to 32 Myr to measure the ``young'' stellar population provides a compromise between time resolution and smoothing over the large uncertainties involved in stellar population modeling (see \citealt{GonzalezDelgado:16} for further discussion).

\subsubsection{Molecular gas surface density}\label{sec:sigmol}

The molecular gas surface density $\Sigma_{\rm mol}$, including H$_2$ and associated helium, is derived by multiplying the CO surface brightness with a Galactic value for the CO-to-H$_2$ conversion factor of \citep[e.g.,][]{Bolatto:13}:
\begin{equation}
\alpha_{\rm CO} = 4.3\;M_\odot\; \rm (K\;km\;s^{-1}\; pc^2)^{-1}\;.
\end{equation}
This column is provided as {\tt sigmol} in the {\tt comom\_xxx} (e.g., {\tt comom\_dil}) table.  As noted in \S\ref{sec:cotable}, an additional column {\tt cosi} contains the multiplicative factor needed to deproject the surface density to face-on assuming a plane-parallel disk.  If used for comparative analysis, the $\cos i$ factor should be used to deproject other surface densities as well.

{The use of a constant CO-to-H$_2$ conversion factor is obviously a simplification, and a number of studies have inferred a decrease of the conversion factor at stellar surface densities of $\gtrsim$100 \Msol\ pc$^{-2}$ \citep[e.g.,][]{Chiang:23} and an increase in the conversion factor at low metallicities \citep[e.g.,][]{Accurso:17}.  We have elected not to implement a variable conversion factor in the current database, for two reasons.  First, the fraction of $\Sigma_* > 100\; \Msol\rm\; pc^{-2}$ points is relatively small ($\sim$20\%) and even above this limit, the inferred dependence on $\Sigma_*$ is relatively weak for the CO(1--0) line ($\propto \Sigma_*^{-0.3}$).  Second, the metallicity range for star-forming spaxels in the EDGE-CALIFA sample is relatively limited (spanning about 0.2 dex or a factor of 1.6) and comparable in size to the systematics in oxygen metallicity calibrations \citep{Marino:13}.  We therefore defer the implementation of a variable conversion factor to a future work that can examine a wider range of conditions.}

\subsubsection{Oxygen abundance}

We determine the gas-phase metallicity by taking a ratio of two emission line ratios, [\OIII]$\lambda$5007/H$\beta$ and [\NII]$\lambda$6583/H$\alpha$ \citep[i.e., the O3N2 method;][]{Alloin:79,Pettini:04}. We use the following prescription from \citet{Marino:13}:
\begin{equation}
12 + {\rm log(O/H)} = 8.533 - 0.214 \ {\rm log} \left( \frac{\rm [\OIII]}{{\rm H}\beta} \times \frac{{\rm H}\alpha}{\rm [\NII]} \right).
\end{equation}
{which has been derived for the logarithm ranging from $-1$ to 1.7, encompassing the range of our data.} Since the metallicity calibration is reliant upon the emission lines arising from \HII\ regions, we also require a ``star-forming'' BPT classification ({\tt SF\_BPT} flag in \S\ref{sec:bpt} below), otherwise the value is blanked.
Error propagation is performed using a standard approach taking into account only the uncertainties in the emission line fluxes and assuming they are independent.
{The resulting values are saved as the {\tt ZOH} column in the {\tt flux\_elines} table (Table~\ref{tab:felines}).}

\subsubsection{BPT Classification of pixels}\label{sec:bpt}

The same line ratios used to estimate metallicity can also provide useful insight into the nature of ionized regions in galaxies.  This is the basis of the most widely used ``BPT'' diagram \citep{1981PASP} which plots [\OIII]$\lambda$5007/H$\beta$ against [\NII]$\lambda$6583/H$\alpha$.  From exploring a wide range of parameters in photoionization models, \citet{Kewley:01} deduced that star-forming regions must lie below a curve defined by the equation
\begin{equation}
\log \left(\frac{[\OIII]}{\rm H\beta}\right) = 1.19 + \frac{0.61}{\log ([\NII]/\rm H\alpha) - 0.47}\;.
\end{equation}
Based on spectra from the Sloan Digital Sky Survey (SDSS), \citet{Kauffmann:03} introduced a more restrictive condition for star-forming regions, which they argued typically lie below the curve defined by
\begin{equation}
\log \left(\frac{[\OIII]}{\rm H\beta}\right) = 1.3 + \frac{0.61}{\log ([\NII]/\rm H\alpha) - 0.05}\;.
\end{equation}
Finally, in the upper right part of the BPT diagram, signifying harder radiation fields, \citet{CidFer:10} find that Seyfert nuclei tend to lie above the line
\begin{equation}
\log \left(\frac{[\OIII]}{\rm H\beta}\right) = 0.48 + 1.01 \log \left(\frac{[\NII]}{\rm H\alpha}\right)\;,
\end{equation}
with points falling below this line characterized as LINERs (although such line ratios are by no means exclusively found in nuclear regions, e.g.\ \citealt{Belfiore:16}).  Following standard convention, we classify points below the \citet{Kauffmann:03} curve as consistent with star-forming (BPT code $-1$), points between the \citet{Kauffmann:03} and \citet{Kewley:01} curves as intermediate or composite (BPT code 0), points above the \citet{Kewley:01} curve and below the \citet{CidFer:10} line as LINER (BPT code 1), and points above the \citet{Kewley:01} curve and above the \citet{CidFer:10} line as Seyfert (BPT code 2).  The BPT code for each sample pixel is included as a column in the {\tt flux\_elines} table, and an additional {\tt SF\_BPT} column is set to True when the BPT code is $-1$ and the H$\alpha$ equivalent width exceeds 6 \AA; we consider such regions to be very likely to be star-forming \citep[e.g.,][]{Barrera:18}.

To assist with interpreting the distribution of points in the diagram, we implemented a probability calculation algorithm for the BPT types. The flux uncertainties for [\NII], [\OIII], H$\alpha$, and H$\beta$ are used to calculate the probability that a datapoint actually lies within the region of the BPT diagram that it is assigned to. This calculation is done by constructing a bivariate Gaussian error distribution assuming independent errors, and numerically estimating the fractional area of the Gaussian that lies in the same BPT region as the center of the Gaussian.  In practice, this is done by sampling points in the distribution using a 5 $\times$ 5 grid spanning $\pm$1$\sigma$ around the nominal value.  Panels (b) and (c) of Figure~\ref{fig:bpt_comp} show how a simple signal-to-noise ratio (S/N) cut applied to the four emission lines compares with imposing the condition that the BPT probability exceed $0.9$. A probability criterion allows more uncertain points to be considered as long as they lie safely within their designated BPT region.  On the other hand, a probability criterion does reject points that lie close to the demarcation lines or where the BPT regions themselves span a narrow extent, since the resulting classification is deemed more uncertain.

\begin{figure*}
    \centerline{\includegraphics[width=0.9\textwidth]{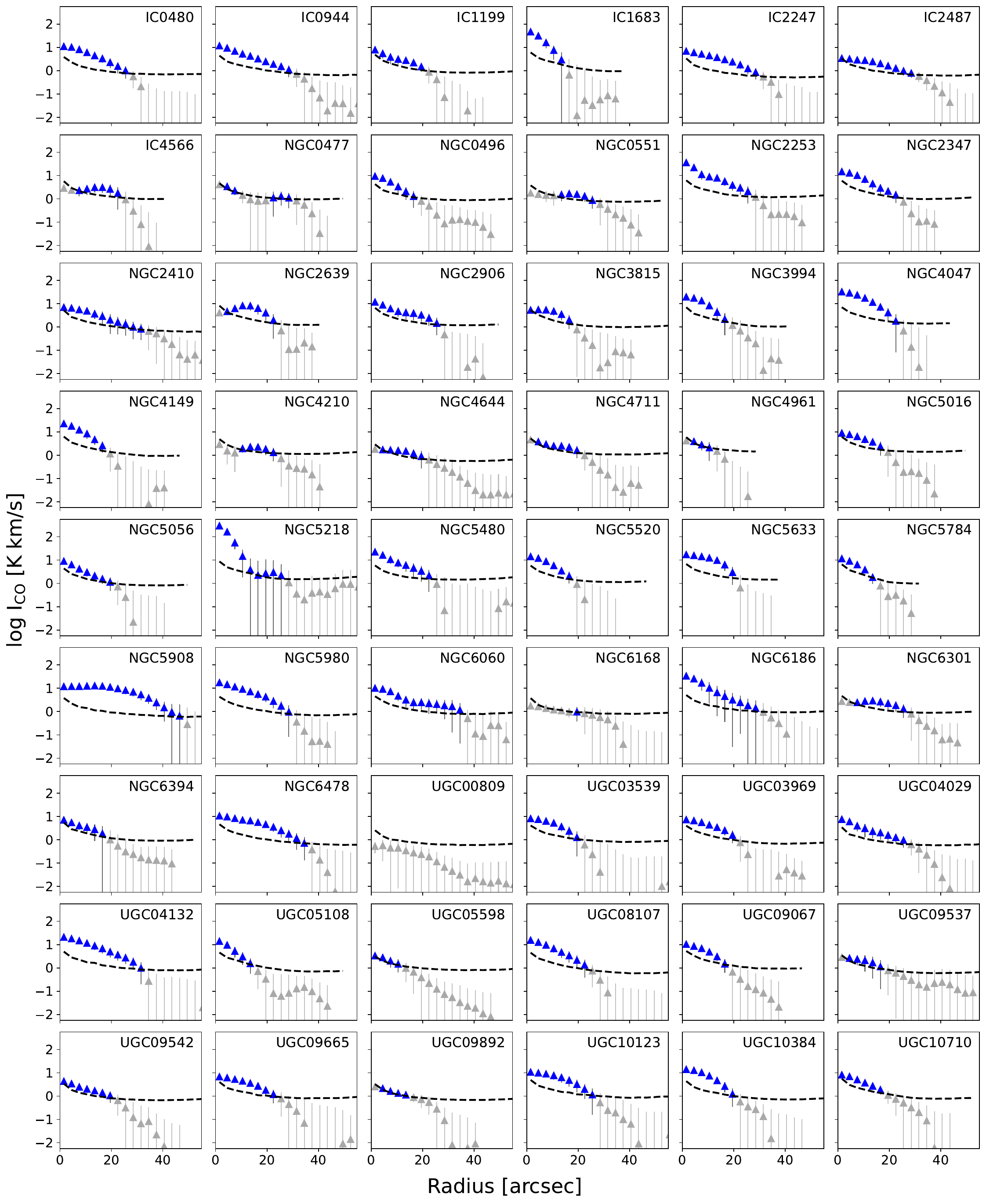}}
    \caption{Inclination-corrected radial CO profiles for the sample of 54 galaxies studied by \citet{Leung:18}.  The dashed curve in each panel represents a 3$\sigma$ detection limit adopting a default velocity width of 200 \kms\ but using the actual mask width where it exceeds 200 \kms\ in width.  Gray points denote radii where the azimuthal average is measured (i.e.\ there are $>$10\% unmasked CO pixels at these radii) but falls below the detection limit and is thus unreliable.} 
    \label{fig:profiles}
\end{figure*}

\begin{figure*}
    \centerline{\includegraphics[width=0.9\textwidth]{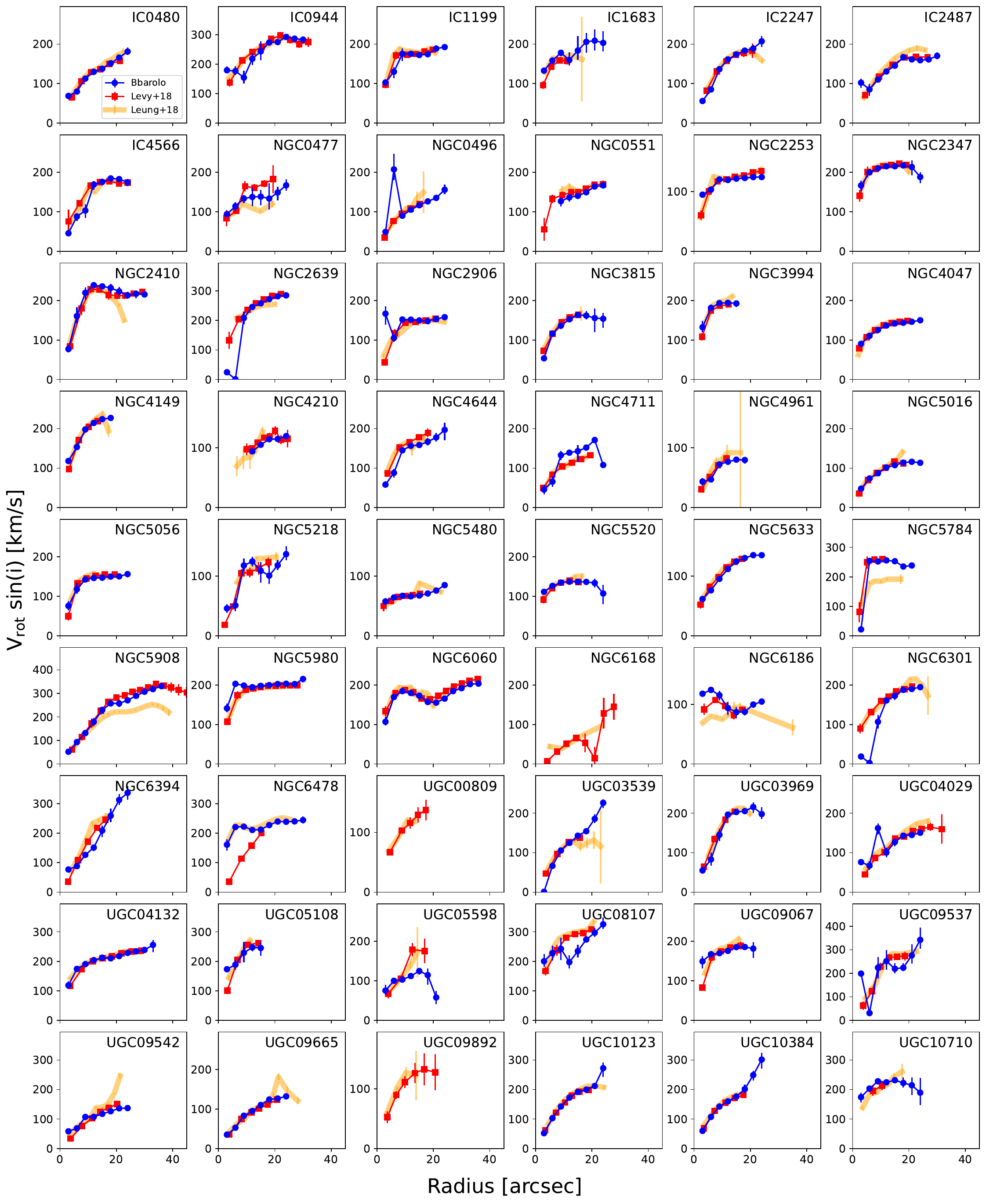}}
    \caption{CO rotation curves derived from the {\tt 3D-Barolo} package ({\it blue circles}) compared to those previously published by \citet{Leung:18} and \citet{Levy:18}, for the sample of 54 galaxies studied by \citet{Leung:18}.  Circular velocities have been multiplied by a $\sin i$ factor to compensate for different adopted inclinations; the plotted curves should thus approximate the observed velocity widths.  Differences are most apparent in the central rings, where varying treatment of beam smearing and velocity dispersion can have significant effects.}
    \label{fig:rotcurves}
\end{figure*}

\begin{figure*}
    \includegraphics[width=\textwidth]{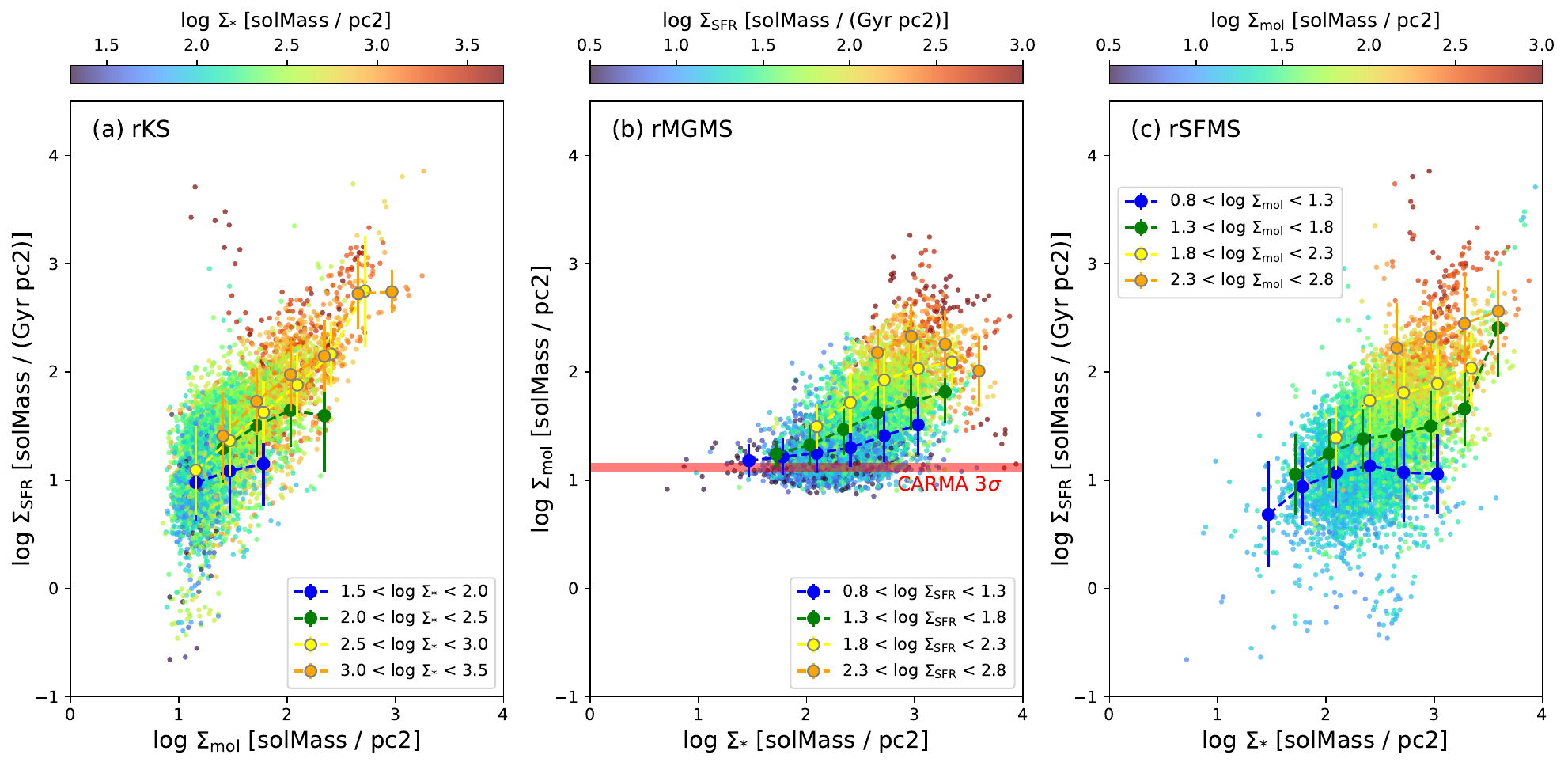}
    \caption{Mutual correlations between molecular gas surface density ($\Sigma_{\rm mol}$), star formation rate surface density (based on extinction-corrected H$\alpha$ emission, $\Sigma_{\rm SFR}$), and stellar surface density ($\Sigma_*$) for the full set of square grid downsampled pixels.  The three panels correspond to the Kennicutt-Schmidt, molecular gas main sequence, and star-forming main sequence diagrams respectively.  Points in each panel are color coded and binned by the value of the third parameter.  A horizontal line in the middle panel denotes the approximate sensitivity limit for the CARMA CO data.}
    \label{fig:threeviews}
\end{figure*}

\begin{figure*}
    \includegraphics[width=\textwidth]{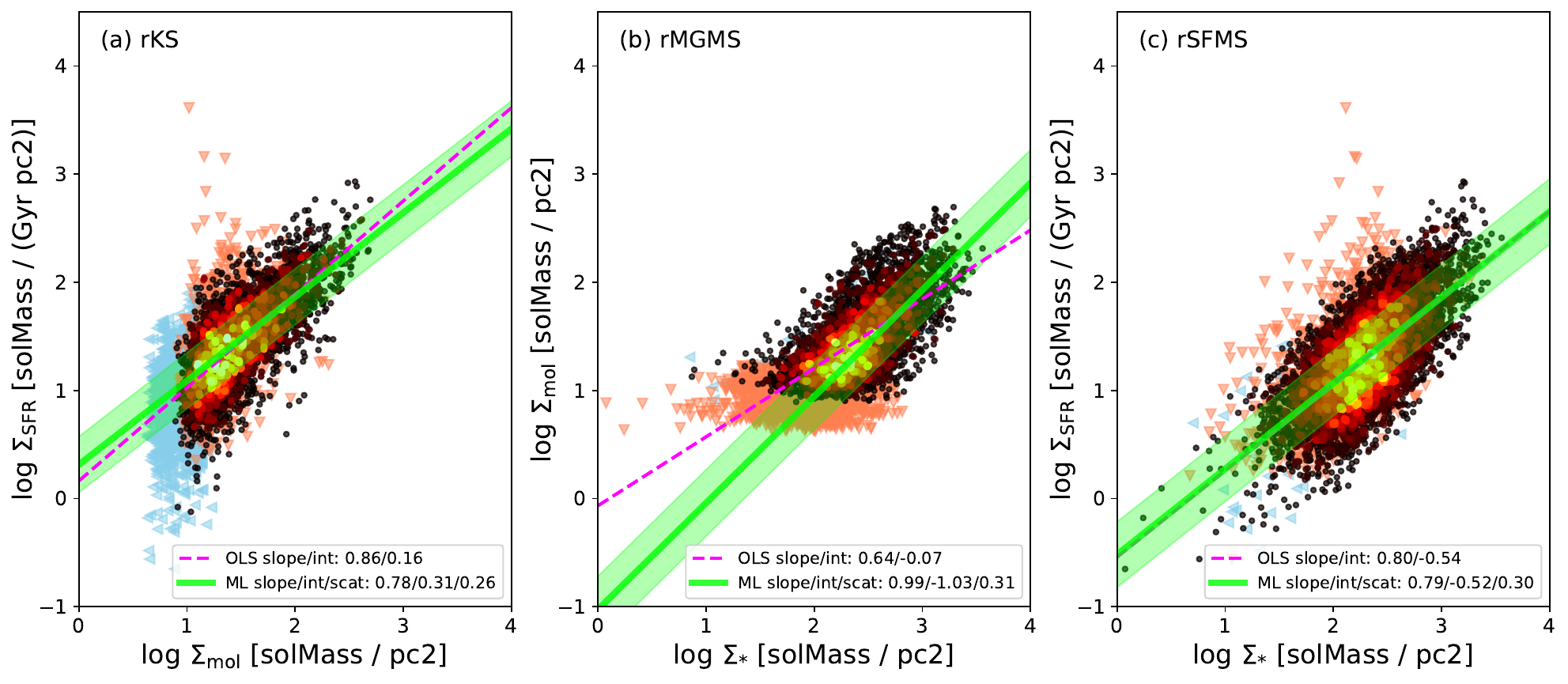}
    \caption{Same panels and data as in Figure~\ref{fig:threeviews}, but limited to pixels classified as star-forming [below the \citealt{Kauffmann:03} BPT curve and with  EW(H$\alpha$) $>$ 6\AA].  Upper limits are shown as pink triangles for the vertical axis and blue triangles for the horizontal axis.  Overlaid are best-fit power law relations based on ordinary least-squares (magenta dotted line) and LEO-Py, which takes into account non-detections (green solid line with vertical scatter indicated by shading).}
    \label{fig:leopy}
\end{figure*}

\begin{figure}
    \includegraphics[width=\columnwidth]{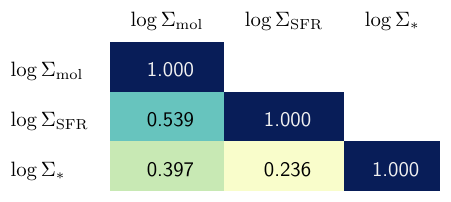}
    \caption{Spearman partial correlation coefficients calculated based on the LEO-Py fits.  The coefficients indicate that the relation between $\Sigma_{\rm SFR}$ and $\Sigma_{\rm mol}$ is the tightest among the three correlation pairs.}
    \label{fig:pcc}
\end{figure}

\subsection{Azimuthally aggregated data}

The 1-D radial profile data are obtained from analysis of the original FITS images and provided as plain text tables.  These include radial CO profiles derived from the ``smoothed'' mask moment-0 maps, as well as circular velocity curves derived from the analysis of \citet{Leung:18} for CO and \citet{Levy:18} for CO and H$\alpha$.  The CO radial profiles and rotation curves are plotted for the subset of 54 EDGE galaxies studied by \citet{Leung:18} in Figures~\ref{fig:profiles} and \ref{fig:rotcurves} respectively.  

Radial CO brightness profiles (deprojected to face-on) are calculated for galaxies with inclinations $<85\arcdeg$ by averaging in concentric rings after setting masked pixels to zero; only rings with 10\% or greater valid pixels are tabulated.  The less restrictive ``smoothed'' moment masks are used in order to include more pixels in the averaging.  The ring widths used are 2\arcsec\ for the native resolution cubes and 3\arcsec\ for the common resolution (7\arcsec) cubes.  Error bars on the points represent the r.m.s.\ deprojected value within the ring for unmasked pixels.  A 3$\sigma$ detection limit profile, derived from collapsing and azimuthally averaging the r.m.s.\ noise cube over a 200 km/s velocity width (or using the actual mask width if larger) is shown as a black dashed line in Fig.~\ref{fig:profiles}; this is higher near the galaxy center where fewer independent beams contribute to the azimuthal average.  Note that the radial profiles may underestimate the true brightness profiles because faint CO emission may still be present at the masked pixels.  Simple alternatives, such as not employing any masking, yield poorer results due to negative residuals in the CO cubes.  A more sensitive approach is to use the H$\alpha$ velocity field to align and stack CO spectra, as presented by \citet{Villanueva:21}.  To derive a radial intensity profile, the stacked spectra are integrated over a velocity window that becomes narrower with increasing galactocentric radius; this is equivalent to using the H$\alpha$ velocity to define the centroid of a signal mask and the galactocentric radius to specify the mask's velocity width.

Rotation curves are also deprojected to represent circular velocities $v_c$, although we plot $v_c \sin i$ in Fig.~\ref{fig:rotcurves} to allow more straightforward comparison of analyses where different inclination angles have been assumed.  In addition to providing the curves previously published by \citet{Leung:18} and \citet{Levy:18}, we have derived new CO rotation curves for 79 galaxies using the {\tt 3D-Barolo} software \citep{DiTeodoro:15} applied to 10 \kms\ channel cubes with the primary beam pattern normalized out (so the noise does not rise toward the edge of the field).  The fits were performed with the center position and disk inclination fixed to their LEDA values and in two passes where the first pass determines an average position angle and systemic velocity across all available rings and the second pass determines the circular speed and gas velocity dispersion (assumed isotropic) of each ring.  {The latter quantity, $\sigma_v(R)$, is not typically available from analysis of velocity images alone.}
The database provides fit results for both passes as separate ECSV files, and provides global orientation parameters for each galaxy as an additional table {\tt edge\_bbpars\_smo7.csv}.
While {\tt 3D-Barolo} implicitly takes beam smearing into account by convolving its model with the synthesized beam before comparison with the data, it {can have} difficulty distinguishing {between} a steeply rising rotation curve near the nucleus {and an increase in} velocity dispersion there, so the {sharp} rise in circular speed inferred near the centers of {some} galaxies may be spurious.
{Overall}, Fig.~\ref{fig:rotcurves} suggests that the {\tt 3D-Barolo} results generally agree with the circular velocity estimates derived by \citet{Leung:18} and \citet{Levy:18}.

\begin{figure*}
    \includegraphics[height = 4in]{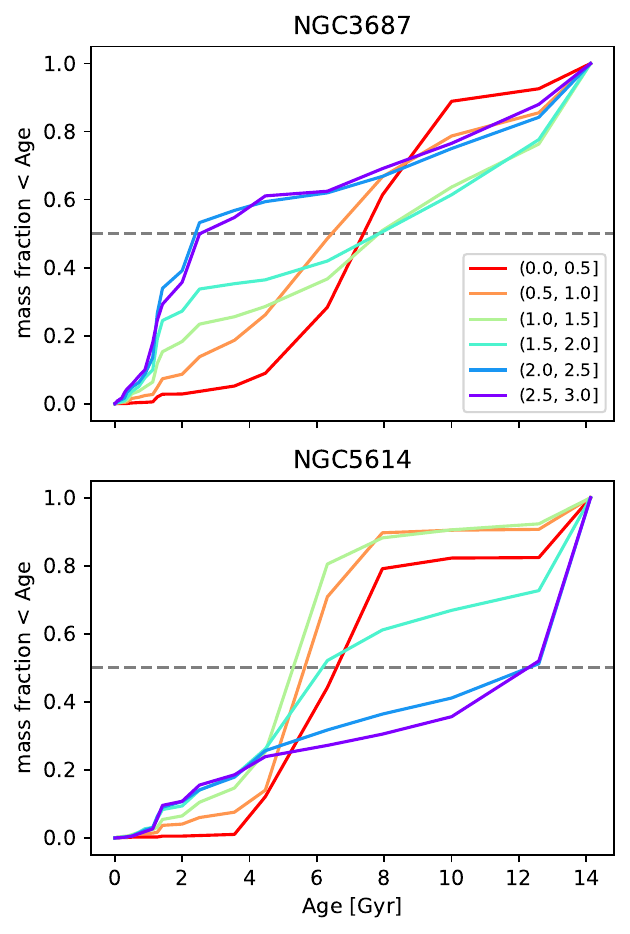}
    \includegraphics[height = 4in]{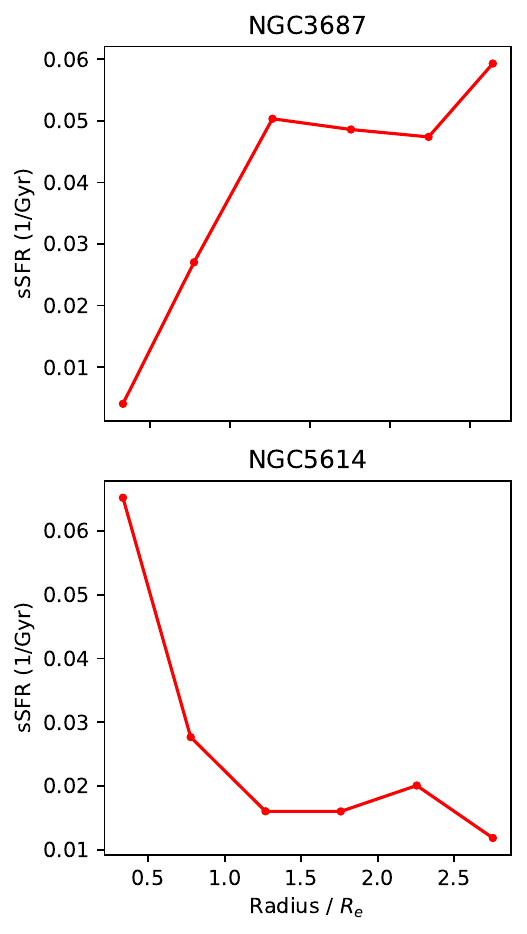}
    \centering
    \caption{Cumulative star formation history, binned in radial rings from 0 to 3\,$R_e$, for NGC 3687 ({\it upper left}, also shown in Fig.~\ref{fig:dotplot}) and NGC 5614 ({\it lower left}).  The intersection of each curve with the dashed line indicates the median age at each radius.  A relatively young age for the outskirts of NGC 3687 is inferred, whereas the inner regions appear younger in NGC 5614.  These general trends are corroborated by the radial profiles of specific SFR ($\Sigma_{\rm SFR}/\Sigma_*$) shown in the right panels.}
    \label{fig:sfhmass}
\end{figure*}

\section{Applications of the database}\label{sec:applications}

\subsection{Star formation scaling relations}\label{sec:scaling}

A straightforward application of the database is to make scatter plots comparing two flux-derived quantities.  Recently there has been much attention given to the relative tightness of observed relations between the stellar surface density $\Sigma_*$, the star formation surface density $\Sigma_{\rm SFR}$, and the molecular gas density $\Sigma_{\rm mol}$ \citep[e.g.,][]{Lin:19,Sanchez:21}.  Figure~\ref{fig:threeviews} shows the mutual correlations between these three variables in EDGE.\@  In each subpanel, the scatter points are colored and vertically binned by the value of the third variable.  While each pair of variables is well correlated, the correlation between $\Sigma_{\rm SFR}$ and $\Sigma_{\rm mol}$, i.e.\ the resolved Kennicutt-Schmidt relation (rKS; \citealt{Kennicutt:98}) appears to be tightest and most independent of the third variable, {as evidenced by the vertical color gradient in panels (b) and (c) of Fig.~\ref{fig:threeviews}}.  The resolved molecular gas main sequence (rMGMS; \citealt{Lin:19}) and resolved star-forming main sequence (rSFMS; \citealt{CanoDiaz:16}) show a much clearer dependence on the third variable.  This result is in agreement with studies {based on the ALMaQUEST survey} that have found the  $\Sigma_{\rm SFR}$--$\Sigma_{\rm mol}$ relation to be the strongest among the three relations \citep{Lin:19,Ellison:20,Baker:22}.

Figure~\ref{fig:leopy} provides a more quantitative characterization of the three relations, making use of the LEO-Py code \citep{Feldmann:19} to deduce the best-fit power law relations taking into account censoring due to non-detections. Only the points from Figure~\ref{fig:threeviews} that are classified as star-forming ({\tt SF\_BPT}=1) are used in the analysis.  LEO-Py derives the fit parameters using likelihood maximization while assuming that the errors in both axes are uncorrelated and each follow a normal distribution.  Non-detections are provided as 3$\sigma$ upper limits.  The best-fit slope, intercept, and vertical scatter are included in the legend of each sub-panel.  The slopes of the three relations range from 0.8 to 1, thus being close to or slightly sub-linear.
For comparison, the results of an ordinary least squares (OLS) fit to the detections only is shown as a dashed magenta line.  The LEO-Py and OLS fits are most discrepant in panel (b), where the large number of CO non-detections leads to a bias towards high values of $\Sigma_{\rm mol}$ at low $\Sigma_*$ and a resultant flattening of the OLS slope.  This is less of an issue in panel (a) because the distribution of $\Sigma_{\rm SFR}$ values at a given $\Sigma_{\rm mol}$ are generally well-sampled and not strongly biased by non-detections.

Figure~\ref{fig:pcc} summarizes the tightness of the three relations by comparing the Spearman partial correlation coefficients between each pair of variables.
For each pair of variables $X$ and $Y$, the partial correlation is the correlation between the residuals $\theta_X$ and $\theta_Y$ after removing the best-fit power laws relating $X$ with $Z$ and $Y$ with $Z$ respectively.  
It therefore signifies how well $X$ and $Y$ are expected to be correlated when $Z$ is held constant.
The partial correlation coefficients confirm that the rKS relation is the tightest correlation between any pair of variables.
Although less tight than the rKS relation, the rMGMS shows a notably stronger partial correlation compared to the rSFMS, in agreement with the conclusion by \citet{Lin:19} {from the ALMaQUEST survey} that the rSFMS is largely a consequence of the other two relations.
{On the other hand, we find that the partial correlation coefficient between \sigsfr\ and \sigstar\ is still positive, whereas \citet{Baker:22} find a value consistent with zero for the ALMaQUEST sample, which includes a substantial fraction ($\sim$50\%) of galaxies below the main sequence.  Whether sample selection effects, in particular the bias towards higher SFRs in EDGE, is responsible for this difference will be an important avenue for future investigation, since it has implications for whether \sigstar\ serves as an important secondary variable in the star formation law \cite[e.g.,][]{Shi:18}.}

\subsection{Spatially resolved star formation histories}

Pipe3D models the stellar light as a linear combination of simple stellar populations taken from the GSD156 library \citep{CidFer:13}, which comprises 156 templates for 39 stellar ages (ranging from 1 Myr to 14.1 Gyr) and four metallicities ($Z/Z_\odot = 0.2$, 0.4, 1, and 1.5).  For each spaxel, the light fraction contributed by each template is recorded in the {\tt SFH} table.  In order to reconstruct the star formation history, we scale the light fraction by each template's mass-to-light ratio, based on a \citet{Salpeter:55} IMF, then multiply the derived mass fraction (summed over metallicity) by the total stellar mass in the spaxel.  We display cumulative star formation histories for two EDGE galaxies, NGC 3687 (which has the prominent H$\alpha$ ring shown in Figure~\ref{fig:dotplot}), and NGC 5614 (an interacting galaxy in the Arp 178 triplet) in Figure~\ref{fig:sfhmass}, along with the corresponding radial profiles of specific SFR (sSFR=$\Sigma_{\rm SFR}/\Sigma_*$) based on extinction-corrected H$\alpha$ emission.  Despite the assumptions involved in recovering the SFH, the radial age trends implied by the SFH profiles are in qualitative agreement with the sSFR profiles, which show high sSFR values (indicative of a recent buildup of stellar mass) at large radii in NGC 3687 and at small radii in NGC 5614.  We note that the ionized gas kinematics in NGC 5614 are highly disturbed \citep{Barrera:15}, consistent with a recent inflow of gas towards the central regions that was triggered by the interaction.

\section{Comparison of Python and SQL databases}\label{sec:compare}

\begin{figure}[t]
    \includegraphics[width = \columnwidth]{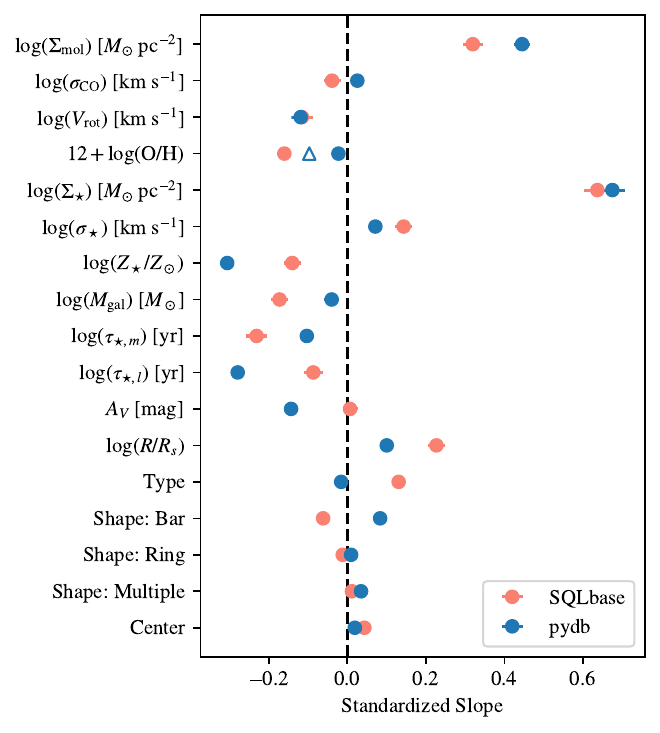}
    \centering
    \caption{Comparison of SQL and {\tt edge\_pydb} databases, following the analysis of \citet{Dey:19}.  The ``standardized slope'' on the horizontal axis is the coefficient of each quantity considered in the multi-linear fit for $\log \Sigma_{\rm SFR}$, scaled by the standard deviation of the same quantity in order to reflect its contribution to the fit.  In both analyses $\Sigma_*$ and $\Sigma_{\rm mol}$ are the dominant contributors to the SFR.  The blue open triangle for O/H shows the {\tt edge\_pydb} result if the N2 estimator used in the SQL database were to be used.}
    \label{fig:sql_pydb}
\end{figure}

As discussed in \S\ref{sec:intro}, a previous implementation of the EDGE database using the SQLite language was used for results presented in \citet{Bolatto:17} and \citet{Dey:19}.  In Figure~\ref{fig:sql_pydb} we compare the results obtained by \citet{Dey:19}, studying the variables that most influence the spatially resolved SFR, with a re-analysis conducted using {\tt edge\_pydb}.  
Displayed is the standardized slope, which represents the coefficient (slope) of the multi-linear fit for a given variable, scaled by its standard deviation to compensate for the fact that variables spanning larger dynamic range {can still exert a strong influence even with a relatively shallow power-law dependence}. 
As expected, $\Sigma_*$ and $\Sigma_{\rm mol}$ remain the most important factors determining $\Sigma_{\rm SFR}$, consistent with the results of \citet{Dey:19}.  The greater importance of $\Sigma_*$ compared to $\Sigma_{\rm mol}$ may come as a surprise given the relative tightness of the resolved K-S relation discussed in \S\ref{sec:scaling}, but note that the limited dynamic range of the CO observations results in the $\Sigma_{\rm SFR}$--$\Sigma_*$ relation spanning a greater logarithmic range in terms of significant detections (cf.\ Figure~\ref{fig:leopy}).  
{Analyses such as those in Figure~\ref{fig:sql_pydb} that do not take the distribution of non-detections into account will therefore favor quantities with larger dynamic range as explanatory variables.}

On the other hand, many of the quantities derived from the stellar analysis (metallicity $Z_*$, age $\tau_*$, extinction $A_V$) show noticeably different standardized slopes compared to the previous work.  An important distinction is that the SQL database was derived from a Pipe3D analysis of the native resolution data cubes from CALIFA.  Smoothing to match the CARMA resolution occurred {\it after} the extraction of the stellar population properties.  On the other hand, for {\tt edge\_pydb} we have degraded the spatial resolution of the CALIFA data to 7\arcsec\ {\it before} running Pipe3D.  Especially age and extinction are likely to vary spatially on scales comparable to the native CALIFA resolution of 2\farcs5 ($\sim$1 kpc), so smoothing an age or extinction map may yield different results compared with sampling a map of the same quantity derived from a smoothed IFU cube.  For both databases, the gas metallicity is derived from smoothed emission-line fluxes, so better consistency would be expected; however, \citet{Dey:19} use the N2 estimator for $12 + \log\rm(O/H)$ \citep{Marino:13} whereas the present database employs the O3N2 estimator.  The open triangle in Figure~\ref{fig:sql_pydb} shows the {\tt edge\_pydb} result if one applies the N2 estimator, which is more consistent with the earlier results.
{Furthermore, the additional smoothing when calculating the Balmer decrement described in \S\ref{sec:sfrdens} was not used in the SQL database---this has the potential to introduce additional shifts in all of the relations involving $\Sigma_{\rm SFR}$.}

\section{Summary and Future Work}\label{sec:summary}

In this paper we have presented our basic approach to compiling a database of galaxy properties derived from spatially resolved CO and IFU observations.  A strong emphasis on usability is essential to our approach, as the database is produced in order to conduct the science rather than as a benefit to the community after the science has been conducted.  We take advantage of self-documenting formats now provided by Astropy tables, and provide sample notebooks that demonstrate how scientifically interesting plots can be generated from the data.

While the EDGE-CALIFA project began with a CALIFA subsample observed with CARMA, it has expanded as additional time has been secured to observe CALIFA galaxies in CO with other millimeter facilities, including the Atacama Pathfinder Experiment (APEX; see \citealt{Colombo:20}), the Atacama Compact Array (ACA; see \citealt{Villanueva:23}), the Large Millimeter Telescope (LMT/GTM), and the Green Bank Telescope (GBT).  The vast majority of CALIFA galaxies still lack spatially resolved CO observations, so the database will continue to grow over time.  In addition, spatially resolved \HI\ observations have recently been obtained for CALIFA galaxies using the Jansky Very Large Array and the Giant Metrewave Radio Telescope (PIs V. Kalinova \& D. Colombo).  The use of the {\tt edge\_carma} prefix in many of the HDF5 file names implicitly recognizes that the current database is tailored to the initial CARMA sample.  Readers wishing to reproduce the exact results of this paper should access the appropriate version of the database as stored in the Github version history.  The EDGE database architecture can easily be adapted to other IFU surveys, particularly when Pipe3D has been used for spectral fitting, and will continue to be updated as Pipe3D itself undergoes further development \citep[e.g.,][]{Sanchez:23}.

The current version of {\tt edge\_pydb} at the time of this publication is indexed on Zenodo as \url{zenodo.org/doi/10.5281/zenodo.10431797}.  The full pixel tables are also available to download from Zenodo at \url{zenodo.org/doi/10.5281/zenodo.10256731}.

\bigskip

We thank the anonymous referee for helpful suggestions which substantially improved this paper.
This work was funded in part by NSF AAG grants 1616199 and 2307440 to the University of Illinois, 1615960 and 2307441 to the University of Maryland, and 1616924 to the University of California-Berkeley.
R.C.L. acknowledges partial support from a NSF Astronomy and Astrophysics Postdoctoral Fellowship under award AST-2102625.
M.R. wishes to acknowledge partial support from ANID Basal FB210003.
V.V. acknowledges support from the scholarship ANID-FULBRIGHT BIO 2016-56160020 and funding from NRAO Student Observing Support SOSPADA-015.
J.B-B acknowledges support from the grant IA-101522 (DGAPA-PAPIIT, UNAM) and funding from the CONACYT grant CF19-39578.
Support for CARMA construction was derived from the Gordon and Betty Moore Foundation, the Kenneth T. and Eileen L. Norris Foundation, the James S. McDonnell Foundation, the Associates of the California Institute of Technology, the University of Chicago, the states of California, Illinois, and Maryland, and the NSF. CARMA development and operations were supported by the NSF under a cooperative agreement and by the CARMA partner universities. This research is based on observations collected at the Centro Astron\'{o}mico Hispano-Alem\'{a}n (CAHA) at Calar Alto, operated jointly by the Max-Planck Institut f\"{u}r Astronomie (MPA) and the Instituto de Astrofisica de Andalucia (CSIC). 
This research has made use of NASA's Astrophysics Data System.
{This publication makes use of data products from the Wide-field Infrared Survey Explorer, which is a joint project of the University of California, Los Angeles, and the Jet Propulsion Laboratory/California Institute of Technology, funded by the National Aeronautics and Space Administration.}


\software{Astropy \citep{Astropy:13, Astropy:18, Astropy:22}, {\tt 3D-Barolo} \citep{DiTeodoro:15}, {\tt LEO-Py} \citep{Feldmann:19}.}

\bibliography{edgebase}{}

\begin{thebibliography}{}
\expandafter\ifx\csname natexlab\endcsname\relax\def\natexlab#1{#1}\fi
\providecommand{\url}[1]{\href{#1}{#1}}
\providecommand{\dodoi}[1]{doi:~\href{http://doi.org/#1}{\nolinkurl{#1}}}
\providecommand{\doeprint}[1]{\href{http://ascl.net/#1}{\nolinkurl{http://ascl.net/#1}}}
\providecommand{\doarXiv}[1]{\href{https://arxiv.org/abs/#1}{\nolinkurl{https://arxiv.org/abs/#1}}}

\bibitem[{{Abazajian} {et~al.}(2009){Abazajian}, {Adelman-McCarthy},
  {Ag{\"u}eros}, {Allam}, {Allende Prieto}, {An}, {Anderson}, {Anderson},
  {Annis}, {Bahcall}, {Bailer-Jones}, {Barentine}, {Bassett}, {Becker},
  {Beers}, {Bell}, {Belokurov}, {Berlind}, {Berman}, {Bernardi}, {Bickerton},
  {Bizyaev}, {Blakeslee}, {Blanton}, {Bochanski}, {Boroski}, {Brewington},
  {Brinchmann}, {Brinkmann}, {Brunner}, {Budav{\'a}ri}, {Carey}, {Carliles},
  {Carr}, {Castander}, {Cinabro}, {Connolly}, {Csabai}, {Cunha}, {Czarapata},
  {Davenport}, {de Haas}, {Dilday}, {Doi}, {Eisenstein}, {Evans}, {Evans},
  {Fan}, {Friedman}, {Frieman}, {Fukugita}, {G{\"a}nsicke}, {Gates},
  {Gillespie}, {Gilmore}, {Gonzalez}, {Gonzalez}, {Grebel}, {Gunn},
  {Gy{\"o}ry}, {Hall}, {Harding}, {Harris}, {Harvanek}, {Hawley}, {Hayes},
  {Heckman}, {Hendry}, {Hennessy}, {Hindsley}, {Hoblitt}, {Hogan}, {Hogg},
  {Holtzman}, {Hyde}, {Ichikawa}, {Ichikawa}, {Im}, {Ivezi{\'c}}, {Jester},
  {Jiang}, {Johnson}, {Jorgensen}, {Juri{\'c}}, {Kent}, {Kessler}, {Kleinman},
  {Knapp}, {Konishi}, {Kron}, {Krzesinski}, {Kuropatkin}, {Lampeitl},
  {Lebedeva}, {Lee}, {Lee}, {French Leger}, {L{\'e}pine}, {Li}, {Lima}, {Lin},
  {Long}, {Loomis}, {Loveday}, {Lupton}, {Magnier}, {Malanushenko},
  {Malanushenko}, {Mandelbaum}, {Margon}, {Marriner}, {Mart{\'\i}nez-Delgado},
  {Matsubara}, {McGehee}, {McKay}, {Meiksin}, {Morrison}, {Mullally}, {Munn},
  {Murphy}, {Nash}, {Nebot}, {Neilsen}, {Newberg}, {Newman}, {Nichol},
  {Nicinski}, {Nieto-Santisteban}, {Nitta}, {Okamura}, {Oravetz}, {Ostriker},
  {Owen}, {Padmanabhan}, {Pan}, {Park}, {Pauls}, {Peoples}, {Percival}, {Pier},
  {Pope}, {Pourbaix}, {Price}, {Purger}, {Quinn}, {Raddick}, {Re Fiorentin},
  {Richards}, {Richmond}, {Riess}, {Rix}, {Rockosi}, {Sako}, {Schlegel},
  {Schneider}, {Scholz}, {Schreiber}, {Schwope}, {Seljak}, {Sesar}, {Sheldon},
  {Shimasaku}, {Sibley}, {Simmons}, {Sivarani}, {Allyn Smith}, {Smith},
  {Smol{\v{c}}i{\'c}}, {Snedden}, {Stebbins}, {Steinmetz}, {Stoughton},
  {Strauss}, {SubbaRao}, {Suto}, {Szalay}, {Szapudi}, {Szkody}, {Tanaka},
  {Tegmark}, {Teodoro}, {Thakar}, {Tremonti}, {Tucker}, {Uomoto}, {Vanden
  Berk}, {Vandenberg}, {Vidrih}, {Vogeley}, {Voges}, {Vogt}, {Wadadekar},
  {Watters}, {Weinberg}, {West}, {White}, {Wilhite}, {Wonders}, {Yanny},
  {Yocum}, {York}, {Zehavi}, {Zibetti}, \& {Zucker}}]{SDSS7}
{Abazajian}, K.~N., {Adelman-McCarthy}, J.~K., {Ag{\"u}eros}, M.~A., {et~al.}
  2009, \apjs, 182, 543, \dodoi{10.1088/0067-0049/182/2/543}

\bibitem[{{Accurso} {et~al.}(2017){Accurso}, {Saintonge}, {Catinella},
  {Cortese}, {Dav{\'e}}, {Dunsheath}, {Genzel}, {Gracia-Carpio}, {Heckman},
  {Jimmy}, {Kramer}, {Li}, {Lutz}, {Schiminovich}, {Schuster}, {Sternberg},
  {Sturm}, {Tacconi}, {Tran}, \& {Wang}}]{Accurso:17}
{Accurso}, G., {Saintonge}, A., {Catinella}, B., {et~al.} 2017, \mnras, 470,
  4750, \dodoi{10.1093/mnras/stx1556}

\bibitem[{{Alloin} {et~al.}(1979){Alloin}, {Collin-Souffrin}, {Joly}, \&
  {Vigroux}}]{Alloin:79}
{Alloin}, D., {Collin-Souffrin}, S., {Joly}, M., \& {Vigroux}, L. 1979, \aap,
  78, 200

\bibitem[{{Astropy Collaboration} {et~al.}(2013){Astropy Collaboration},
  {Robitaille}, {Tollerud}, {Greenfield}, {Droettboom}, {Bray}, {Aldcroft},
  {Davis}, {Ginsburg}, {Price-Whelan}, {Kerzendorf}, {Conley}, {Crighton},
  {Barbary}, {Muna}, {Ferguson}, {Grollier}, {Parikh}, {Nair}, {Unther},
  {Deil}, {Woillez}, {Conseil}, {Kramer}, {Turner}, {Singer}, {Fox}, {Weaver},
  {Zabalza}, {Edwards}, {Azalee Bostroem}, {Burke}, {Casey}, {Crawford},
  {Dencheva}, {Ely}, {Jenness}, {Labrie}, {Lim}, {Pierfederici}, {Pontzen},
  {Ptak}, {Refsdal}, {Servillat}, \& {Streicher}}]{Astropy:13}
{Astropy Collaboration}, {Robitaille}, T.~P., {Tollerud}, E.~J., {et~al.} 2013,
  \aap, 558, A33, \dodoi{10.1051/0004-6361/201322068}

\bibitem[{{Astropy Collaboration} {et~al.}(2018){Astropy Collaboration},
  {Price-Whelan}, {Sip{\H{o}}cz}, {G{\"u}nther}, {Lim}, {Crawford}, {Conseil},
  {Shupe}, {Craig}, {Dencheva}, {Ginsburg}, {Vand erPlas}, {Bradley},
  {P{\'e}rez-Su{\'a}rez}, {de Val-Borro}, {Aldcroft}, {Cruz}, {Robitaille},
  {Tollerud}, {Ardelean}, {Babej}, {Bach}, {Bachetti}, {Bakanov}, {Bamford},
  {Barentsen}, {Barmby}, {Baumbach}, {Berry}, {Biscani}, {Boquien}, {Bostroem},
  {Bouma}, {Brammer}, {Bray}, {Breytenbach}, {Buddelmeijer}, {Burke},
  {Calderone}, {Cano Rodr{\'\i}guez}, {Cara}, {Cardoso}, {Cheedella}, {Copin},
  {Corrales}, {Crichton}, {D'Avella}, {Deil}, {Depagne}, {Dietrich}, {Donath},
  {Droettboom}, {Earl}, {Erben}, {Fabbro}, {Ferreira}, {Finethy}, {Fox},
  {Garrison}, {Gibbons}, {Goldstein}, {Gommers}, {Greco}, {Greenfield},
  {Groener}, {Grollier}, {Hagen}, {Hirst}, {Homeier}, {Horton}, {Hosseinzadeh},
  {Hu}, {Hunkeler}, {Ivezi{\'c}}, {Jain}, {Jenness}, {Kanarek}, {Kendrew},
  {Kern}, {Kerzendorf}, {Khvalko}, {King}, {Kirkby}, {Kulkarni}, {Kumar},
  {Lee}, {Lenz}, {Littlefair}, {Ma}, {Macleod}, {Mastropietro}, {McCully},
  {Montagnac}, {Morris}, {Mueller}, {Mumford}, {Muna}, {Murphy}, {Nelson},
  {Nguyen}, {Ninan}, {N{\"o}the}, {Ogaz}, {Oh}, {Parejko}, {Parley}, {Pascual},
  {Patil}, {Patil}, {Plunkett}, {Prochaska}, {Rastogi}, {Reddy Janga},
  {Sabater}, {Sakurikar}, {Seifert}, {Sherbert}, {Sherwood-Taylor}, {Shih},
  {Sick}, {Silbiger}, {Singanamalla}, {Singer}, {Sladen}, {Sooley},
  {Sornarajah}, {Streicher}, {Teuben}, {Thomas}, {Tremblay}, {Turner},
  {Terr{\'o}n}, {van Kerkwijk}, {de la Vega}, {Watkins}, {Weaver}, {Whitmore},
  {Woillez}, {Zabalza}, \& {Astropy Contributors}}]{Astropy:18}
{Astropy Collaboration}, {Price-Whelan}, A.~M., {Sip{\H{o}}cz}, B.~M., {et~al.}
  2018, \aj, 156, 123, \dodoi{10.3847/1538-3881/aabc4f}

\bibitem[{{Astropy Collaboration} {et~al.}(2022){Astropy Collaboration},
  {Price-Whelan}, {Lim}, {Earl}, {Starkman}, {Bradley}, {Shupe}, {Patil},
  {Corrales}, {Brasseur}, {N{"o}the}, {Donath}, {Tollerud}, {Morris},
  {Ginsburg}, {Vaher}, {Weaver}, {Tocknell}, {Jamieson}, {van Kerkwijk},
  {Robitaille}, {Merry}, {Bachetti}, {G{"u}nther}, {Aldcroft},
  {Alvarado-Montes}, {Archibald}, {B{'o}di}, {Bapat}, {Barentsen}, {Baz{'a}n},
  {Biswas}, {Boquien}, {Burke}, {Cara}, {Cara}, {Conroy}, {Conseil}, {Craig},
  {Cross}, {Cruz}, {D'Eugenio}, {Dencheva}, {Devillepoix}, {Dietrich},
  {Eigenbrot}, {Erben}, {Ferreira}, {Foreman-Mackey}, {Fox}, {Freij}, {Garg},
  {Geda}, {Glattly}, {Gondhalekar}, {Gordon}, {Grant}, {Greenfield}, {Groener},
  {Guest}, {Gurovich}, {Handberg}, {Hart}, {Hatfield-Dodds}, {Homeier},
  {Hosseinzadeh}, {Jenness}, {Jones}, {Joseph}, {Kalmbach}, {Karamehmetoglu},
  {Ka{l}uszy{'n}ski}, {Kelley}, {Kern}, {Kerzendorf}, {Koch}, {Kulumani},
  {Lee}, {Ly}, {Ma}, {MacBride}, {Maljaars}, {Muna}, {Murphy}, {Norman},
  {O'Steen}, {Oman}, {Pacifici}, {Pascual}, {Pascual-Granado}, {Patil},
  {Perren}, {Pickering}, {Rastogi}, {Roulston}, {Ryan}, {Rykoff}, {Sabater},
  {Sakurikar}, {Salgado}, {Sanghi}, {Saunders}, {Savchenko}, {Schwardt},
  {Seifert-Eckert}, {Shih}, {Jain}, {Shukla}, {Sick}, {Simpson},
  {Singanamalla}, {Singer}, {Singhal}, {Sinha}, {Sip{H{o}}cz}, {Spitler},
  {Stansby}, {Streicher}, {{{S}}umak}, {Swinbank}, {Taranu}, {Tewary},
  {Tremblay}, {Val-Borro}, {Van Kooten}, {Vasovi{'c}}, {Verma}, {de Miranda
  Cardoso}, {Williams}, {Wilson}, {Winkel}, {Wood-Vasey}, {Xue}, {Yoachim},
  {Zhang}, {Zonca}, \& {Astropy Project Contributors}}]{Astropy:22}
{Astropy Collaboration}, {Price-Whelan}, A.~M., {Lim}, P.~L., {et~al.} 2022,
  apj, 935, 167, \dodoi{10.3847/1538-4357/ac7c74}

\bibitem[{{Baker} {et~al.}(2022){Baker}, {Maiolino}, {Bluck}, {Lin}, {Ellison},
  {Belfiore}, {Pan}, \& {Thorp}}]{Baker:22}
{Baker}, W.~M., {Maiolino}, R., {Bluck}, A. F.~L., {et~al.} 2022, \mnras, 510,
  3622, \dodoi{10.1093/mnras/stab3672}

\bibitem[{{Baldwin} {et~al.}(1981){Baldwin}, {Phillips}, \&
  {Terlevich}}]{1981PASP}
{Baldwin}, J.~A., {Phillips}, M.~M., \& {Terlevich}, R. 1981, \pasp, 93, 5,
  \dodoi{10.1086/130766}

\bibitem[{{Barrera-Ballesteros} {et~al.}(2015){Barrera-Ballesteros},
  {Garc{\'\i}a-Lorenzo}, {Falc{\'o}n-Barroso}, {van de Ven}, {Lyubenova},
  {Wild}, {M{\'e}ndez-Abreu}, {S{\'a}nchez}, {Marquez}, {Masegosa},
  {Monreal-Ibero}, {Ziegler}, {del Olmo}, {Verdes-Montenegro},
  {Garc{\'\i}a-Benito}, {Husemann}, {Mast}, {Kehrig}, {Iglesias-Paramo},
  {Marino}, {Aguerri}, {Walcher}, {V{\'\i}lchez}, {Bomans}, {Cortijo-Ferrero},
  {Gonz{\'a}lez Delgado}, {Bland-Hawthorn}, {McIntosh}, \&
  {Bekerait{\.{e}}}}]{Barrera:15}
{Barrera-Ballesteros}, J.~K., {Garc{\'\i}a-Lorenzo}, B., {Falc{\'o}n-Barroso},
  J., {et~al.} 2015, \aap, 582, A21, \dodoi{10.1051/0004-6361/201424935}

\bibitem[{{Barrera-Ballesteros} {et~al.}(2018){Barrera-Ballesteros}, {Heckman},
  {S{\'a}nchez}, {Zakamska}, {Cleary}, {Zhu}, {Brinkmann}, {Drory}, \& {THE
  MaNGA TEAM}}]{Barrera:18}
{Barrera-Ballesteros}, J.~K., {Heckman}, T., {S{\'a}nchez}, S.~F., {et~al.}
  2018, \apj, 852, 74, \dodoi{10.3847/1538-4357/aa9b31}

\bibitem[{{Barrera-Ballesteros} {et~al.}(2021){Barrera-Ballesteros},
  {S{\'a}nchez}, {Heckman}, {Wong}, {Bolatto}, {Ostriker}, {Rosolowsky},
  {Carigi}, {Vogel}, {Levy}, {Colombo}, {Luo}, \& {Cao}}]{Barrera:21}
{Barrera-Ballesteros}, J.~K., {S{\'a}nchez}, S.~F., {Heckman}, T., {et~al.}
  2021, \mnras, 503, 3643, \dodoi{10.1093/mnras/stab755}

\bibitem[{{Belfiore} {et~al.}(2016){Belfiore}, {Maiolino}, {Maraston},
  {Emsellem}, {Bershady}, {Masters}, {Yan}, {Bizyaev}, {Boquien}, {Brownstein},
  {Bundy}, {Drory}, {Heckman}, {Law}, {Roman-Lopes}, {Pan}, {Stanghellini},
  {Thomas}, {Weijmans}, \& {Westfall}}]{Belfiore:16}
{Belfiore}, F., {Maiolino}, R., {Maraston}, C., {et~al.} 2016, \mnras, 461,
  3111, \dodoi{10.1093/mnras/stw1234}

\bibitem[{{Bigiel} {et~al.}(2008){Bigiel}, {Leroy}, {Walter}, {Brinks}, {de
  Blok}, {Madore}, \& {Thornley}}]{Bigiel:08}
{Bigiel}, F., {Leroy}, A., {Walter}, F., {et~al.} 2008, \aj, 136, 2846,
  \dodoi{10.1088/0004-6256/136/6/2846}

\bibitem[{{Bitsakis} {et~al.}(2019){Bitsakis}, {S{\'a}nchez}, {Ciesla},
  {Bonfini}, {Charmandaris}, {Cervantes Sodi}, {Maragkoudakis}, {Diaz-Santos},
  \& {Zezas}}]{Bitsakis:19}
{Bitsakis}, T., {S{\'a}nchez}, S.~F., {Ciesla}, L., {et~al.} 2019, \mnras, 483,
  370, \dodoi{10.1093/mnras/sty2857}

\bibitem[{{Bolatto} {et~al.}(2013){Bolatto}, {Wolfire}, \&
  {Leroy}}]{Bolatto:13}
{Bolatto}, A.~D., {Wolfire}, M., \& {Leroy}, A.~K. 2013, \araa, 51, 207,
  \dodoi{10.1146/annurev-astro-082812-140944}

\bibitem[{{Bolatto} {et~al.}(2017){Bolatto}, {Wong}, {Utomo}, {Blitz}, {Vogel},
  {S{\'a}nchez}, {Barrera-Ballesteros}, {Cao}, {Colombo}, {Dannerbauer},
  {Garc{\'\i}a-Benito}, {Herrera-Camus}, {Husemann}, {Kalinova}, {Leroy},
  {Leung}, {Levy}, {Mast}, {Ostriker}, {Rosolowsky}, {Sandstrom}, {Teuben},
  {van de Ven}, \& {Walter}}]{Bolatto:17}
{Bolatto}, A.~D., {Wong}, T., {Utomo}, D., {et~al.} 2017, \apj, 846, 159,
  \dodoi{10.3847/1538-4357/aa86aa}

\bibitem[{{Bottinelli} {et~al.}(1983){Bottinelli}, {Gouguenheim}, {Paturel}, \&
  {de Vaucouleurs}}]{Bottinelli:83}
{Bottinelli}, L., {Gouguenheim}, L., {Paturel}, G., \& {de Vaucouleurs}, G.
  1983, \aap, 118, 4

\bibitem[{{Brinchmann} {et~al.}(2004){Brinchmann}, {Charlot}, {White},
  {Tremonti}, {Kauffmann}, {Heckman}, \& {Brinkmann}}]{Brinchmann:04}
{Brinchmann}, J., {Charlot}, S., {White}, S.~D.~M., {et~al.} 2004, \mnras, 351,
  1151, \dodoi{10.1111/j.1365-2966.2004.07881.x}

\bibitem[{{Calzetti} {et~al.}(2010){Calzetti}, {Wu}, {Hong}, {Kennicutt},
  {Lee}, {Dale}, {Engelbracht}, {van Zee}, {Draine}, {Hao}, {Gordon},
  {Moustakas}, {Murphy}, {Regan}, {Begum}, {Block}, {Dalcanton}, {Funes}, {Gil
  de Paz}, {Johnson}, {Sakai}, {Skillman}, {Walter}, {Weisz}, {Williams}, \&
  {Wu}}]{Calzetti:10}
{Calzetti}, D., {Wu}, S.~Y., {Hong}, S., {et~al.} 2010, \apj, 714, 1256,
  \dodoi{10.1088/0004-637X/714/2/1256}

\bibitem[{{Cano-D{\'\i}az} {et~al.}(2016){Cano-D{\'\i}az}, {S{\'a}nchez},
  {Zibetti}, {Ascasibar}, {Bland-Hawthorn}, {Ziegler}, {Gonz{\'a}lez Delgado},
  {Walcher}, {Garc{\'\i}a-Benito}, {Mast}, {Mendoza-P{\'e}rez},
  {Falc{\'o}n-Barroso}, {Galbany}, {Husemann}, {Kehrig}, {Marino},
  {S{\'a}nchez-Bl{\'a}zquez}, {L{\'o}pez-Cob{\'a}}, {L{\'o}pez-S{\'a}nchez}, \&
  {Vilchez}}]{CanoDiaz:16}
{Cano-D{\'\i}az}, M., {S{\'a}nchez}, S.~F., {Zibetti}, S., {et~al.} 2016,
  \apjl, 821, L26, \dodoi{10.3847/2041-8205/821/2/L26}

\bibitem[{{Cao} {et~al.}(2023){Cao}, {Wong}, {Bolatto}, {Leroy}, {Rosolowsky},
  {Utomo}, {S{\'a}nchez}, {Barrera-Ballesteros}, {Levy}, {Colombo}, {Blitz},
  {Vogel}, {Puschnig}, {Villanueva}, \& {Rubio}}]{Cao:23}
{Cao}, Y., {Wong}, T., {Bolatto}, A.~D., {et~al.} 2023, \apjs, 268, 3,
  \dodoi{10.3847/1538-4365/acd840}

\bibitem[{{Cardelli} {et~al.}(1989){Cardelli}, {Clayton}, \&
  {Mathis}}]{Cardelli:89}
{Cardelli}, J.~A., {Clayton}, G.~C., \& {Mathis}, J.~S. 1989, \apj, 345, 245,
  \dodoi{10.1086/167900}

\bibitem[{{Catal{\'a}n-Torrecilla} {et~al.}(2015){Catal{\'a}n-Torrecilla}, {Gil
  de Paz}, {Castillo-Morales}, {Iglesias-P{\'a}ramo}, {S{\'a}nchez},
  {Kennicutt}, {P{\'e}rez-Gonz{\'a}lez}, {Marino}, {Walcher}, {Husemann},
  {Garc{\'\i}a-Benito}, {Mast}, {Gonz{\'a}lez Delgado}, {Mu{\~n}oz-Mateos},
  {Bland-Hawthorn}, {Bomans}, {Del Olmo}, {Galbany}, {Gomes}, {Kehrig},
  {L{\'o}pez-S{\'a}nchez}, {Mendoza}, {Monreal-Ibero}, {P{\'e}rez-Torres},
  {S{\'a}nchez-Bl{\'a}zquez}, {Vilchez}, \& {Califa
  Collaboration}}]{Catalan:15}
{Catal{\'a}n-Torrecilla}, C., {Gil de Paz}, A., {Castillo-Morales}, A.,
  {et~al.} 2015, \aap, 584, A87, \dodoi{10.1051/0004-6361/201526023}

\bibitem[{{Chiang} {et~al.}(2023){Chiang}, {Sandstrom}, {Chastenet}, {Bolatto},
  {Koch}, {Leroy}, {Sun}, {Teng}, \& {Williams}}]{Chiang:23}
{Chiang}, I.-D., {Sandstrom}, K.~M., {Chastenet}, J., {et~al.} 2023, arXiv
  e-prints, arXiv:2311.00407, \dodoi{10.48550/arXiv.2311.00407}

\bibitem[{{Cid Fernandes} {et~al.}(2010){Cid Fernandes}, {Stasi{\'n}ska},
  {Schlickmann}, {Mateus}, {Vale Asari}, {Schoenell}, \&
  {Sodr{\'e}}}]{CidFer:10}
{Cid Fernandes}, R., {Stasi{\'n}ska}, G., {Schlickmann}, M.~S., {et~al.} 2010,
  \mnras, 403, 1036, \dodoi{10.1111/j.1365-2966.2009.16185.x}

\bibitem[{{Cid Fernandes} {et~al.}(2013){Cid Fernandes}, {P{\'e}rez},
  {Garc{\'\i}a Benito}, {Gonz{\'a}lez Delgado}, {de Amorim}, {S{\'a}nchez},
  {Husemann}, {Falc{\'o}n Barroso}, {S{\'a}nchez-Bl{\'a}zquez}, {Walcher}, \&
  {Mast}}]{CidFer:13}
{Cid Fernandes}, R., {P{\'e}rez}, E., {Garc{\'\i}a Benito}, R., {et~al.} 2013,
  \aap, 557, A86, \dodoi{10.1051/0004-6361/201220616}

\bibitem[{{Colombo} {et~al.}(2018){Colombo}, {Kalinova}, {Utomo}, {Rosolowsky},
  {Bolatto}, {Levy}, {Wong}, {Sanchez}, {Leroy}, {Ostriker}, {Blitz}, {Vogel},
  {Mast}, {Garc{\'\i}a-Benito}, {Husemann}, {Dannerbauer}, {Ellmeier}, \&
  {Cao}}]{Colombo:18}
{Colombo}, D., {Kalinova}, V., {Utomo}, D., {et~al.} 2018, \mnras, 475, 1791,
  \dodoi{10.1093/mnras/stx3233}

\bibitem[{{Colombo} {et~al.}(2020){Colombo}, {Sanchez}, {Bolatto}, {Kalinova},
  {Wei{\ss}}, {Wong}, {Rosolowsky}, {Vogel}, {Barrera-Ballesteros},
  {Dannerbauer}, {Cao}, {Levy}, {Utomo}, \& {Blitz}}]{Colombo:20}
{Colombo}, D., {Sanchez}, S.~F., {Bolatto}, A.~D., {et~al.} 2020, \aap, 644,
  A97, \dodoi{10.1051/0004-6361/202039005}

\bibitem[{{Dey} {et~al.}(2019){Dey}, {Rosolowsky}, {Cao}, {Bolatto}, {Sanchez},
  {Utomo}, {Colombo}, {Kalinova}, {Wong}, {Blitz}, {Vogel}, {Loeppky}, \&
  {Garc{\'\i}a-Benito}}]{Dey:19}
{Dey}, B., {Rosolowsky}, E., {Cao}, Y., {et~al.} 2019, \mnras, 488, 1926,
  \dodoi{10.1093/mnras/stz1777}

\bibitem[{{Di Teodoro} \& {Fraternali}(2015)}]{DiTeodoro:15}
{Di Teodoro}, E.~M., \& {Fraternali}, F. 2015, \mnras, 451, 3021,
  \dodoi{10.1093/mnras/stv1213}

\bibitem[{{Ellison} {et~al.}(2020){Ellison}, {Thorp}, {Lin}, {Pan}, {Bluck},
  {Scudder}, {Teimoorinia}, {S{\'a}nchez}, \& {Sargent}}]{Ellison:20}
{Ellison}, S.~L., {Thorp}, M.~D., {Lin}, L., {et~al.} 2020, \mnras, 493, L39,
  \dodoi{10.1093/mnrasl/slz179}

\bibitem[{{Ellison} {et~al.}(2021){Ellison}, {Wong}, {S{\'a}nchez}, {Colombo},
  {Bolatto}, {Barrera-Ballesteros}, {Garc{\'\i}a-Benito}, {Kalinova}, {Luo},
  {Rubio}, \& {Vogel}}]{Ellison:21}
{Ellison}, S.~L., {Wong}, T., {S{\'a}nchez}, S.~F., {et~al.} 2021, \mnras, 505,
  L46, \dodoi{10.1093/mnrasl/slab047}

\bibitem[{{Feldmann}(2019)}]{Feldmann:19}
{Feldmann}, R. 2019, Astronomy and Computing, 29, 100331,
  \dodoi{10.1016/j.ascom.2019.100331}

\bibitem[{{Garay-Solis} {et~al.}(2023){Garay-Solis}, {Barrera-Ballesteros},
  {Colombo}, {S{\'a}nchez}, {Lugo-Aranda}, {Villanueva}, {Wong}, \&
  {Bolatto}}]{GaraySolis:23}
{Garay-Solis}, Y., {Barrera-Ballesteros}, J.~K., {Colombo}, D., {et~al.} 2023,
  \apj, 952, 122, \dodoi{10.3847/1538-4357/acd781}

\bibitem[{{Gonz{\'a}lez Delgado} {et~al.}(2016){Gonz{\'a}lez Delgado}, {Cid
  Fernandes}, {P{\'e}rez}, {Garc{\'\i}a-Benito}, {L{\'o}pez Fern{\'a}ndez},
  {Lacerda}, {Cortijo-Ferrero}, {de Amorim}, {Vale Asari}, {S{\'a}nchez},
  {Walcher}, {Wisotzki}, {Mast}, {Alves}, {Ascasibar}, {Bland-Hawthorn},
  {Galbany}, {Kennicutt}, {M{\'a}rquez}, {Masegosa}, {Moll{\'a}},
  {S{\'a}nchez-Bl{\'a}zquez}, \& {V{\'\i}lchez}}]{GonzalezDelgado:16}
{Gonz{\'a}lez Delgado}, R.~M., {Cid Fernandes}, R., {P{\'e}rez}, E., {et~al.}
  2016, \aap, 590, A44, \dodoi{10.1051/0004-6361/201628174}

\bibitem[{{Kauffmann} {et~al.}(2003){Kauffmann}, {Heckman}, {Tremonti},
  {Brinchmann}, {Charlot}, {White}, {Ridgway}, {Brinkmann}, {Fukugita}, {Hall},
  {Ivezi{\'c}}, {Richards}, \& {Schneider}}]{Kauffmann:03}
{Kauffmann}, G., {Heckman}, T.~M., {Tremonti}, C., {et~al.} 2003, \mnras, 346,
  1055, \dodoi{10.1111/j.1365-2966.2003.07154.x}

\bibitem[{{Kennicutt}(1998)}]{Kennicutt:98}
{Kennicutt}, Robert~C., J. 1998, \apj, 498, 541, \dodoi{10.1086/305588}

\bibitem[{{Kennicutt} \& {Evans}(2012)}]{Kennicutt:12}
{Kennicutt}, R.~C., \& {Evans}, N.~J. 2012, \araa, 50, 531,
  \dodoi{10.1146/annurev-astro-081811-125610}

\bibitem[{{Kewley} {et~al.}(2001){Kewley}, {Dopita}, {Sutherland}, {Heisler},
  \& {Trevena}}]{Kewley:01}
{Kewley}, L.~J., {Dopita}, M.~A., {Sutherland}, R.~S., {Heisler}, C.~A., \&
  {Trevena}, J. 2001, \apj, 556, 121, \dodoi{10.1086/321545}

\bibitem[{{Kroupa} {et~al.}(1993){Kroupa}, {Tout}, \& {Gilmore}}]{Kroupa:93}
{Kroupa}, P., {Tout}, C.~A., \& {Gilmore}, G. 1993, \mnras, 262, 545,
  \dodoi{10.1093/mnras/262.3.545}

\bibitem[{{Leung} {et~al.}(2018){Leung}, {Leaman}, {van de Ven}, {Lyubenova},
  {Zhu}, {Bolatto}, {Falc{\'o}n-Barroso}, {Blitz}, {Dannerbauer}, {Fisher},
  {Levy}, {Sanchez}, {Utomo}, {Vogel}, {Wong}, \& {Ziegler}}]{Leung:18}
{Leung}, G. Y.~C., {Leaman}, R., {van de Ven}, G., {et~al.} 2018, \mnras, 477,
  254, \dodoi{10.1093/mnras/sty288}

\bibitem[{{Levy} {et~al.}(2018){Levy}, {Bolatto}, {Teuben}, {S{\'a}nchez},
  {Barrera-Ballesteros}, {Blitz}, {Colombo}, {Garc{\'\i}a-Benito},
  {Herrera-Camus}, {Husemann}, {Kalinova}, {Lan}, {Leung}, {Mast}, {Utomo},
  {van de Ven}, {Vogel}, \& {Wong}}]{Levy:18}
{Levy}, R.~C., {Bolatto}, A.~D., {Teuben}, P., {et~al.} 2018, \apj, 860, 92,
  \dodoi{10.3847/1538-4357/aac2e5}

\bibitem[{{Levy} {et~al.}(2019){Levy}, {Bolatto}, {S{\'a}nchez}, {Blitz},
  {Colombo}, {Kalinova}, {L{\'o}pez-Cob{\'a}}, {Ostriker}, {Teuben}, {Utomo},
  {Vogel}, \& {Wong}}]{Levy:19}
{Levy}, R.~C., {Bolatto}, A.~D., {S{\'a}nchez}, S.~F., {et~al.} 2019, \apj,
  882, 84, \dodoi{10.3847/1538-4357/ab2ed4}

\bibitem[{{Lin} {et~al.}(2019){Lin}, {Pan}, {Ellison}, {Belfiore}, {Shi},
  {S{\'a}nchez}, {Hsieh}, {Rowlands}, {Ramya}, {Thorp}, {Li}, \&
  {Maiolino}}]{Lin:19}
{Lin}, L., {Pan}, H.-A., {Ellison}, S.~L., {et~al.} 2019, \apjl, 884, L33,
  \dodoi{10.3847/2041-8213/ab4815}

\bibitem[{{Makarov} {et~al.}(2014){Makarov}, {Prugniel}, {Terekhova},
  {Courtois}, \& {Vauglin}}]{Makarov:14}
{Makarov}, D., {Prugniel}, P., {Terekhova}, N., {Courtois}, H., \& {Vauglin},
  I. 2014, \aap, 570, A13, \dodoi{10.1051/0004-6361/201423496}

\bibitem[{{Marino} {et~al.}(2013){Marino}, {Rosales-Ortega}, {S{\'a}nchez},
  {Gil de Paz}, {V{\'\i}lchez}, {Miralles-Caballero}, {Kehrig},
  {P{\'e}rez-Montero}, {Stanishev}, {Iglesias-P{\'a}ramo}, {D{\'\i}az},
  {Castillo-Morales}, {Kennicutt}, {L{\'o}pez-S{\'a}nchez}, {Galbany},
  {Garc{\'\i}a-Benito}, {Mast}, {Mendez-Abreu}, {Monreal-Ibero}, {Husemann},
  {Walcher}, {Garc{\'\i}a-Lorenzo}, {Masegosa}, {Del Olmo Orozco},
  {Mour{\~a}o}, {Ziegler}, {Moll{\'a}}, {Papaderos},
  {S{\'a}nchez-Bl{\'a}zquez}, {Gonz{\'a}lez Delgado}, {Falc{\'o}n-Barroso},
  {Roth}, {van de Ven}, \& {Califa Team}}]{Marino:13}
{Marino}, R.~A., {Rosales-Ortega}, F.~F., {S{\'a}nchez}, S.~F., {et~al.} 2013,
  \aap, 559, A114, \dodoi{10.1051/0004-6361/201321956}

\bibitem[{{Masters} {et~al.}(2014){Masters}, {Crook}, {Hong}, {Jarrett},
  {Koribalski}, {Macri}, {Springob}, \& {Staveley-Smith}}]{Masters:14}
{Masters}, K.~L., {Crook}, A., {Hong}, T., {et~al.} 2014, \mnras, 443, 1044,
  \dodoi{10.1093/mnras/stu1225}

\bibitem[{{Pettini} \& {Pagel}(2004)}]{Pettini:04}
{Pettini}, M., \& {Pagel}, B. E.~J. 2004, \mnras, 348, L59,
  \dodoi{10.1111/j.1365-2966.2004.07591.x}

\bibitem[{{Saintonge} \& {Catinella}(2022)}]{Saintonge:22}
{Saintonge}, A., \& {Catinella}, B. 2022, \araa, 60, 319,
  \dodoi{10.1146/annurev-astro-021022-043545}

\bibitem[{{Saintonge} {et~al.}(2011){Saintonge}, {Kauffmann}, {Wang}, {Kramer},
  {Tacconi}, {Buchbender}, {Catinella}, {Graci{\'a}-Carpio}, {Cortese},
  {Fabello}, {Fu}, {Genzel}, {Giovanelli}, {Guo}, {Haynes}, {Heckman},
  {Krumholz}, {Lemonias}, {Li}, {Moran}, {Rodriguez-Fernandez}, {Schiminovich},
  {Schuster}, \& {Sievers}}]{Saintonge:11}
{Saintonge}, A., {Kauffmann}, G., {Wang}, J., {et~al.} 2011, \mnras, 415, 61,
  \dodoi{10.1111/j.1365-2966.2011.18823.x}

\bibitem[{{Saintonge} {et~al.}(2017){Saintonge}, {Catinella}, {Tacconi},
  {Kauffmann}, {Genzel}, {Cortese}, {Dav{\'e}}, {Fletcher},
  {Graci{\'a}-Carpio}, {Kramer}, {Heckman}, {Janowiecki}, {Lutz}, {Rosario},
  {Schiminovich}, {Schuster}, {Wang}, {Wuyts}, {Borthakur}, {Lamperti}, \&
  {Roberts-Borsani}}]{Saintonge:17}
{Saintonge}, A., {Catinella}, B., {Tacconi}, L.~J., {et~al.} 2017, ApJS, 233,
  22, \dodoi{10.3847/1538-4365/aa97e0}

\bibitem[{{Salim} {et~al.}(2007){Salim}, {Rich}, {Charlot}, {Brinchmann},
  {Johnson}, {Schiminovich}, {Seibert}, {Mallery}, {Heckman}, {Forster},
  {Friedman}, {Martin}, {Morrissey}, {Neff}, {Small}, {Wyder}, {Bianchi},
  {Donas}, {Lee}, {Madore}, {Milliard}, {Szalay}, {Welsh}, \& {Yi}}]{Salim:07}
{Salim}, S., {Rich}, R.~M., {Charlot}, S., {et~al.} 2007, \apjs, 173, 267,
  \dodoi{10.1086/519218}

\bibitem[{{Salpeter}(1955)}]{Salpeter:55}
{Salpeter}, E.~E. 1955, \apj, 121, 161, \dodoi{10.1086/145971}

\bibitem[{{S{\'a}nchez} {et~al.}(2023){S{\'a}nchez}, {Galbany}, {Walcher},
  {Garc{\'\i}a-Benito}, \& {Barrera-Ballesteros}}]{Sanchez:23}
{S{\'a}nchez}, S.~F., {Galbany}, L., {Walcher}, C.~J., {Garc{\'\i}a-Benito},
  R., \& {Barrera-Ballesteros}, J.~K. 2023, \mnras, 526, 5555,
  \dodoi{10.1093/mnras/stad3119}

\bibitem[{{S{\'a}nchez} {et~al.}(2012){S{\'a}nchez}, {Kennicutt}, {Gil de Paz},
  {van de Ven}, {V{\'\i}lchez}, {Wisotzki}, {Walcher}, {Mast}, {Aguerri},
  {Albiol-P{\'e}rez}, {Alonso-Herrero}, {Alves}, {Bakos}, {Bart{\'a}kov{\'a}},
  {Bland-Hawthorn}, {Boselli}, {Bomans}, {Castillo-Morales}, {Cortijo-Ferrero},
  {de Lorenzo-C{\'a}ceres}, {Del Olmo}, {Dettmar}, {D{\'\i}az}, {Ellis},
  {Falc{\'o}n-Barroso}, {Flores}, {Gallazzi}, {Garc{\'\i}a-Lorenzo},
  {Gonz{\'a}lez Delgado}, {Gruel}, {Haines}, {Hao}, {Husemann},
  {Igl{\'e}sias-P{\'a}ramo}, {Jahnke}, {Johnson}, {Jungwiert}, {Kalinova},
  {Kehrig}, {Kupko}, {L{\'o}pez-S{\'a}nchez}, {Lyubenova}, {Marino},
  {M{\'a}rmol-Queralt{\'o}}, {M{\'a}rquez}, {Masegosa}, {Meidt},
  {Mendez-Abreu}, {Monreal-Ibero}, {Montijo}, {Mour{\~a}o}, {Palacios-Navarro},
  {Papaderos}, {Pasquali}, {Peletier}, {P{\'e}rez}, {P{\'e}rez}, {Quirrenbach},
  {Rela{\~n}o}, {Rosales-Ortega}, {Roth}, {Ruiz-Lara},
  {S{\'a}nchez-Bl{\'a}zquez}, {Sengupta}, {Singh}, {Stanishev}, {Trager},
  {Vazdekis}, {Viironen}, {Wild}, {Zibetti}, \& {Ziegler}}]{Sanchez:12}
{S{\'a}nchez}, S.~F., {Kennicutt}, R.~C., {Gil de Paz}, A., {et~al.} 2012,
  A\&A, 538, A8, \dodoi{10.1051/0004-6361/201117353}

\bibitem[{{S{\'a}nchez} {et~al.}(2016{\natexlab{a}}){S{\'a}nchez},
  {Garc{\'\i}a-Benito}, {Zibetti}, {Walcher}, {Husemann}, {Mendoza}, {Galbany},
  {Falc{\'o}n-Barroso}, {Mast}, {Aceituno}, {Aguerri}, {Alves}, {Amorim},
  {Ascasibar}, {Barrado-Navascues}, {Barrera-Ballesteros}, {Bekerait{\`e}},
  {Bland -Hawthorn}, {Cano D{\'\i}az}, {Cid Fernandes}, {Cavichia}, {Cortijo},
  {Dannerbauer}, {Demleitner}, {D{\'\i}az}, {Dettmar}, {de
  Lorenzo-C{\'a}ceres}, {del Olmo}, {Galazzi}, {Garc{\'\i}a-Lorenzo}, {Gil de
  Paz}, {Gonz{\'a}lez Delgado}, {Holmes}, {Igl{\'e}sias-P{\'a}ramo}, {Kehrig},
  {Kelz}, {Kennicutt}, {Kleemann}, {Lacerda}, {L{\'o}pez Fern{\'a}ndez},
  {L{\'o}pez S{\'a}nchez}, {Lyubenova}, {Marino}, {M{\'a}rquez},
  {Mendez-Abreu}, {Moll{\'a}}, {Monreal-Ibero}, {Ortega Minakata},
  {Torres-Papaqui}, {P{\'e}rez}, {Rosales-Ortega}, {Roth},
  {S{\'a}nchez-Bl{\'a}zquez}, {Schilling}, {Spekkens}, {Vale Asari}, {van den
  Bosch}, {van de Ven}, {Vilchez}, {Wild}, {Wisotzki}, {Y{\i}ld{\i}r{\i}m}, \&
  {Ziegler}}]{Sanchez:16}
{S{\'a}nchez}, S.~F., {Garc{\'\i}a-Benito}, R., {Zibetti}, S., {et~al.}
  2016{\natexlab{a}}, \aap, 594, A36, \dodoi{10.1051/0004-6361/201628661}

\bibitem[{{S{\'a}nchez} {et~al.}(2016{\natexlab{b}}){S{\'a}nchez}, {P{\'e}rez},
  {S{\'a}nchez-Bl{\'a}zquez}, {Garc{\'\i}a-Benito}, {Ibarra-Mede},
  {Gonz{\'a}lez}, {Rosales-Ortega}, {S{\'a}nchez-Menguiano}, {Ascasibar},
  {Bitsakis}, {Law}, {Cano-D{\'\i}az}, {L{\'o}pez-Cob{\'a}}, {Marino}, {Gil de
  Paz}, {L{\'o}pez-S{\'a}nchez}, {Barrera-Ballesteros}, {Galbany}, {Mast},
  {Abril-Melgarejo}, \& {Roman-Lopes}}]{Sanchez:Rx}
{S{\'a}nchez}, S.~F., {P{\'e}rez}, E., {S{\'a}nchez-Bl{\'a}zquez}, P., {et~al.}
  2016{\natexlab{b}}, \rmxaa, 52, 171.
\newblock \doarXiv{1602.01830}

\bibitem[{{S{\'a}nchez} {et~al.}(2021){S{\'a}nchez}, {Barrera-Ballesteros},
  {Colombo}, {Wong}, {Bolatto}, {Rosolowsky}, {Vogel}, {Levy}, {Kalinova},
  {Alvarez-Hurtado}, {Luo}, \& {Cao}}]{Sanchez:21}
{S{\'a}nchez}, S.~F., {Barrera-Ballesteros}, J.~K., {Colombo}, D., {et~al.}
  2021, \mnras, 503, 1615, \dodoi{10.1093/mnras/stab442}

\bibitem[{{Schlegel} {et~al.}(1998){Schlegel}, {Finkbeiner}, \&
  {Davis}}]{Schlegel:98}
{Schlegel}, D.~J., {Finkbeiner}, D.~P., \& {Davis}, M. 1998, \apj, 500, 525,
  \dodoi{10.1086/305772}

\bibitem[{{Shi} {et~al.}(2018){Shi}, {Yan}, {Armus}, {Gu}, {Helou}, {Qiu},
  {Gwyn}, {Stierwalt}, {Fang}, {Chen}, {Zhou}, {Wu}, {Zheng}, {Zhang}, {Gao},
  \& {Wang}}]{Shi:18}
{Shi}, Y., {Yan}, L., {Armus}, L., {et~al.} 2018, \apj, 853, 149,
  \dodoi{10.3847/1538-4357/aaa3e6}

\bibitem[{{Springob} {et~al.}(2005){Springob}, {Haynes}, {Giovanelli}, \&
  {Kent}}]{Springob:05}
{Springob}, C.~M., {Haynes}, M.~P., {Giovanelli}, R., \& {Kent}, B.~R. 2005,
  \apjs, 160, 149, \dodoi{10.1086/431550}

\bibitem[{{Utomo} {et~al.}(2017){Utomo}, {Bolatto}, {Wong}, {Ostriker},
  {Blitz}, {Sanchez}, {Colombo}, {Leroy}, {Cao}, {Dannerbauer},
  {Garcia-Benito}, {Husemann}, {Kalinova}, {Levy}, {Mast}, {Rosolowsky}, \&
  {Vogel}}]{Utomo:17}
{Utomo}, D., {Bolatto}, A.~D., {Wong}, T., {et~al.} 2017, \apj, 849, 26,
  \dodoi{10.3847/1538-4357/aa88c0}

\bibitem[{{van Driel} {et~al.}(2001){van Driel}, {Marcum}, {Gallagher},
  {Wilcots}, {Guidoux}, \& {Monnier Ragaigne}}]{vanDriel:01}
{van Driel}, W., {Marcum}, P., {Gallagher}, J.~S., I., {et~al.} 2001, \aap,
  378, 370, \dodoi{10.1051/0004-6361:20011241}

\bibitem[{{Villanueva} {et~al.}(2021){Villanueva}, {Bolatto}, {Vogel}, {Levy},
  {S{\'a}nchez}, {Barrera-Ballesteros}, {Wong}, {Rosolowsky}, {Colombo},
  {Rubio}, {Cao}, {Kalinova}, {Leroy}, {Utomo}, {Herrera-Camus}, {Blitz}, \&
  {Luo}}]{Villanueva:21}
{Villanueva}, V., {Bolatto}, A., {Vogel}, S., {et~al.} 2021, \apj, 923, 60,
  \dodoi{10.3847/1538-4357/ac2b29}

\bibitem[{{Villanueva} {et~al.}(2023){Villanueva}, {Bolatto}, {Vogel}, {Wong},
  {Leroy}, {Sanchez}, {Levy}, {Rosolowsky}, {Colombo}, {Kalinova}, {Cronin},
  {Teuben}, {Rubio}, \& {Bazzi}}]{Villanueva:23}
{Villanueva}, V., {Bolatto}, A.~D., {Vogel}, S.~N., {et~al.} 2023, arXiv
  e-prints, arXiv:2312.03995, \dodoi{10.48550/arXiv.2312.03995}

\bibitem[{{Wong} \& {Blitz}(2002)}]{Wong:02}
{Wong}, T., \& {Blitz}, L. 2002, \apj, 569, 157, \dodoi{10.1086/339287}

\bibitem[{{Wong} {et~al.}(2013){Wong}, {Xue}, {Bolatto}, {Leroy}, {Blitz},
  {Rosolowsky}, {Bigiel}, {Fisher}, {Ott}, {Rahman}, {Vogel}, \&
  {Walter}}]{Wong:13}
{Wong}, T., {Xue}, R., {Bolatto}, A.~D., {et~al.} 2013, \apjl, 777, L4,
  \dodoi{10.1088/2041-8205/777/1/L4}

\bibitem[{{Wright} {et~al.}(2010){Wright}, {Eisenhardt}, {Mainzer}, {Ressler},
  {Cutri}, {Jarrett}, {Kirkpatrick}, {Padgett}, {McMillan}, {Skrutskie},
  {Stanford}, {Cohen}, {Walker}, {Mather}, {Leisawitz}, {Gautier}, {McLean},
  {Benford}, {Lonsdale}, {Blain}, {Mendez}, {Irace}, {Duval}, {Liu}, {Royer},
  {Heinrichsen}, {Howard}, {Shannon}, {Kendall}, {Walsh}, {Larsen}, {Cardon},
  {Schick}, {Schwalm}, {Abid}, {Fabinsky}, {Naes}, \& {Tsai}}]{Wright:10}
{Wright}, E.~L., {Eisenhardt}, P. R.~M., {Mainzer}, A.~K., {et~al.} 2010, \aj,
  140, 1868, \dodoi{10.1088/0004-6256/140/6/1868}

\end{thebibliography}
\bibliographystyle{aasjournal}

\end{document}